\documentclass[11pt,a4paper]{article}
\pdfoutput=1
\usepackage{jheppub}
\usepackage{rotating}
\usepackage{array}
\usepackage{amsmath}
\usepackage{slashed}
\usepackage{booktabs}
\usepackage[pdftex,table]{xcolor}
\usepackage{mathtools}
\usepackage{empheq}

\newcommand{\vev}{v_\text{EW}}

\newcommand{\be}{\bar{\epsilon}}
\newcommand{\Ep}{E' + \be - E}

\preprint{DESY 18-133}

\title{BBN constraints on MeV-scale dark sectors. \\ Part II. Electromagnetic decays}

\author{Marco Hufnagel,}
\author{Kai Schmidt-Hoberg}
\author{and Sebastian Wild}

\affiliation{DESY, Notkestra\ss e 85, D-22607 Hamburg, Germany}

\emailAdd{marco.hufnagel@desy.de}
\emailAdd{kai.schmidt-hoberg@desy.de}
\emailAdd{sebastian.wild@desy.de}

\abstract{
Meta-stable dark sector particles decaying into electrons or photons may non-trivially change the Hubble rate, lead to entropy injection into the thermal bath of Standard Model particles and may also
photodisintegrate light nuclei formed in the early universe.
We study generic constraints from Big Bang Nucleosynthesis on such a setup, with a particular emphasis on MeV-scale particles which are neither fully relativistic 
nor non-relativistic during all times relevant for Big Bang Nucleosynthesis. 
We apply our results to a simple model of self-interacting dark matter with a light scalar mediator. This setup turns out to be severely constrained by these considerations in
combination with direct dark matter searches and will be fully tested with the next generation of low-threshold direct detection experiments.}
\keywords{}

\begin{document}
\maketitle

\section{Introduction}

It is well known that fundamental particles leave their imprint on cosmological probes such as Big Bang Nucleosynthesis (BBN) or Cosmic Microwave Background (CMB) observations.  
In fact the remarkable agreement between the measured abundances of light elements such as deuterium and helium and the corresponding predictions within the Standard Model (SM) of particle physics implies that 
from about one second after the Big Bang the SM provides a good description of the early cosmological evolution. In particular, any deviations from the SM up to MeV-scale energies are strongly constrained~\cite{2016RvMP...88a5004C,Olive:2016xmw,Shvartsman:1969mm,Steigman:1977kc,Scherrer:1987rr,1988ApJ...331...33S}.

There are various effects which influence the light element abundances. Of particular relevance is the expansion rate $H$ of the Universe during the time of BBN, as it determines at which point in time
protons and neutrons fall out of thermodynamic equilibrium and hence sets the ratio of the corresponding number densities. The Hubble rate, in turn, is fully determined by the total energy density, which receives contributions from all particles, including those beyond the SM. In particular, even fully decoupled dark sectors can be probed via their effect on the expansion rate, a scenario which has been explored in detail recently~\cite{Hufnagel:2017dgo}.
 
In this paper we extend this study to a different class of models, where additional effects from entropy production as well as the destruction of light nuclei via photodisintegration are relevant. 
Specifically we study the effect of exotic particles which are meta-stable and decay into electrons and photons during or after BBN. While previous studies have investigated similar scenarios~\cite{Sarkar:1984tt,Ellis:1984er,Scherrer:1987rr,Kawasaki:1994sc,Cyburt:2002uv,Jedamzik:2006xz,Poulin:2015woa,Poulin:2015opa}, it has always been assumed that
the decaying particles are {\it non-relativistic} during BBN. Here we study the {\it fully general} case without such simplifying assumptions, which turns out to be relevant for a number of phenomenological applications.

To this end, we develop in section~\ref{sec:cosmo_evolution} the formalism for the cosmological evolution of an MeV-scale particle $\phi$ which decays into $e^+ e^-$ and/or $\gamma \gamma$ with a lifetime $10^{-2} \, \text{s} \lesssim \tau_\phi \lesssim 10^8\,\text{s}$. To evaluate the effect on BBN, the calculation of the modified Hubble rate, as well as the non-standard evolution of the baryon-to-photon ratio $\eta$ originating from the production of entropy in the decays of $\phi$ are of particular relevance. In section~\ref{sec:abundance_of_light_elements} we evaluate the impact of the modified Hubble rate as well as the non-standard time-dependence of $\eta$ on the light nuclear abundances using a modified version of \texttt{AlterBBN}~\cite{Arbey:2011nf}.
We then study the additional modifications of those nuclear abundances due to photodisintegration reactions induced by the decay products of $\phi$, which become relevant at $t \gtrsim 10^4\,$s. To this end, we first study the cascade of MeV-scale photons, electrons and positrons interacting with CMB photons as well as the background electrons and nuclei. In this context it is mandatory to properly solve the coupled Boltzmann equations, as the 
often utilised `universal photon spectrum' does not apply for the parameter values of interest as shown in~\cite{Poulin:2015woa, Poulin:2015opa}.
We then compute the time evolution of the abundances of light elements which can be destroyed or created by collisions with those additional photons.

Our general results are obtained by comparing the predicted nuclear abundances to observations for varying particle mass, abundance and lifetime in section~\ref{sec:results}. We also take into account the possibility of a different dark sector temperature and provide additional material in appendix~\ref{app:plot_collection} for convenience. In section~\ref{sec:SIDM} we apply our general results in conjunction with bounds from direct dark matter (DM) searches to a simple model of DM coupled to a light scalar mediator, which features potentially large DM self-interactions and has been extensively studied in the literature~\cite{Buckley:2009in,Loeb:2010gj,Kaplinghat:2013yxa,Kainulainen:2015sva,Kahlhoefer:2017umn}. Finally, in appendix~\ref{app:rates_cascade} we provide the rates for all processes relevant to the cascade of MeV-scale photons, electrons and positrons for reference, correcting a couple of typos found in the literature.


\section{Cosmological evolution of the decaying particle and its decay products}
\label{sec:cosmo_evolution}

Let us consider a scalar or vector particle $\phi$ which decays exclusively into $e^+ e^-$ and/or $\gamma \gamma$ with a lifetime $\tau_\phi$. We assume that $\phi$ has been produced at high temperatures (well before the onset of BBN), e.g.~via freeze-out of a dark matter particle or via the decay of some other heavy state. Defining $f_\phi (t, E)$ to be the phase space distribution function of the particle $\phi$ at time $t$, the initial condition for the cosmological evolution of $\phi$ reads $f_\phi (t_0, E) = f_\phi^{(0)}(E)$, with a function $f_\phi^{(0)}(E)$ that depends on the production history of the particle. For example, if $\phi$ has been produced via freeze-out in the visible sector, $f_\phi^{(0)}(E)$ is simply a Bose-Einstein distribution with temperature $T(t_0)$.

We assume that for $t > t_0$ the particle is fully decoupled\footnote{After decoupling of all process changing the number density of $\phi$, the particle may still be in kinetic equilibrium with the thermal bath. It has however been shown in~\cite{Hufnagel:2017dgo} that in this case there are only negligible deviations from the picture we consider here.}, i.e.~it is only subject to redshift and decay. Its phase space distribution then evolves according to the Boltzmann equation~\cite{Kolb:1990vq}
\begin{align}
E \frac{\partial f_\phi(t,E)}{\partial t} - H(t) (E^2 - m_\phi^2) \frac{\partial f_\phi(t,E)}{\partial E} = - \frac{m_\phi}{\tau_\phi} f_\phi (t,E) \;\,.
\label{eq:fphi_BoltzmannEqn}
\end{align}
Here, $m_\phi$ is the mass of $\phi$ and $H(t)$ the Hubble rate. 
In the following, we will take into account the dependence of $H(t)$ on the additional energy density of $\phi$ as well as on the modified radiation density of the SM sector by recursively calculating the evolution of $f_\phi(t,E)$.
Specifically, we start with the SM expression for $H(t)$ and update its value in each recursion step.

The solution of eq.~(\ref{eq:fphi_BoltzmannEqn}) is given by
\begin{align}
f_\phi(t,E) &= \exp \left( -\frac{1}{\tau_\phi} \int_{t_0}^{t} \text{d} \lambda\; \frac{m_\phi}{E_\star(t,\lambda,E)} \right) \cdot f_\phi^{(0)} \left( \,E_\star(t,t_0,E)\,\right)
\label{eq:fphi_sol}
\end{align}
with
\begin{align}
E_\star(t,\lambda,E) &\equiv \sqrt{m_\phi^2 + (E^2 - m_\phi^2) \exp \left( 2 \int_{\lambda}^t \text{d}x\; H(x)\right)}\;\,.
\end{align}
The interpretation of this expression is simple: For $\tau_\phi \to \infty$, the evolution of $f_\phi(t,E)$ simply follows from redshifting the momenta according to $p \propto 1/R$, leading to the relation between $E_\star$ and the redshifted energy $E$ given above. For finite lifetimes, one also has to take into account the decay of $\phi$ with the appropriate time-averaged Lorentz factor as described by the exponential function. Note that the solution (\ref{eq:fphi_sol}) does not assume anything about $\phi$ being ultra- or non-relativistic.

At redshifts $z \gtrsim 2 \times 10^6$, corresponding to $t \lesssim 2 \times 10^8\,$s, the electromagnetic particles produced in the decay of $\phi$ rapidly thermalise with the background photons~\cite{Chluba:2011hw,Poulin:2016anj}. This process increases the temperature $T(t)$ of the photon bath compared to the value $T^\text{(SM)}(t)$ in the SM, which affects BBN in two possible ways: first, the increased energy density $\rho_\text{SM}(t) \propto T(t)^4$ of the SM degrees of freedom (together with the energy density $\rho_\phi$ of the particle $\phi$) leads to a modified expansion rate during BBN. Secondly, the decay of $\phi$ produces entropy in the visible sector, leading to a non-trivial time-dependence of the baryon-to-photon ratio~$\eta$.

The evolution of the photon temperature $T(t)$ can be deduced from the Friedmann equations applied to the combination of both the dark and the visible sector:
\begin{align}
\dot{\rho}_\text{tot} + 3 H \left(\rho_\text{tot} + p_\text{tot}\right) = 0 \;\,,
\end{align}
with the total energy density $\rho_\text{tot} = \rho_\text{SM} + \rho_\phi$ and the pressure $p_\text{tot} = p_\text{SM} + p_\phi$ being the sum of the contributions from the SM heat bath and the particle $\phi$, respectively. Integrating the Boltzmann equation~(\ref{eq:fphi_BoltzmannEqn}) over all energies gives $\dot{\rho}_\phi + 3 H(\rho_\phi + p_\phi) = - m_\phi n_\phi/\tau_\phi$, where $n_\phi(t)$ is the number density of $\phi$ following from eq.~(\ref{eq:fphi_sol}). The energy density and pressure of the SM heat bath thus evolve according to
\begin{align}
\dot{\rho}_\text{SM} + 3 H \left(\rho_\text{SM} + p_\text{SM}\right) = \frac{m_\phi}{\tau_\phi} n_\phi \;\,.
\label{eq:rhoR_evolutionequation}
\end{align}
The decay products of $\phi$ only directly heat up the electromagnetically interacting particles (photons, electrons and positrons), but not necessarily the neutrinos. Hence, we have to solve eq.~(\ref{eq:rhoR_evolutionequation}) separately for $t < t_{\nu\text{-dec}}$ and $t > t_{\nu\text{-dec}}$, where $t_{\nu\text{-dec}}$ is the time of neutrino decoupling. In the SM this is given by $T^\text{(SM)}(t_{\nu\text{-dec}}^{\text{(SM)}}) \simeq 1.4\,$MeV~\cite{Dolgov:2002wy}. As the rates for neutrino interactions scale as $T^5$, we estimate the time of neutrino decoupling in presence of a non-standard time-temperature relationship $T(t)$ via\footnote{We assume that neutrino decoupling occurs instantaneously at $t_{\nu\text{-dec}}$. We refer to~\cite{Kawasaki:1999na} for a more detailed treatment of neutrino decoupling, relevant in the context of a scenario with a matter-dominated epoch prior to BBN.}
\begin{equation}
T(t_{\nu\text{-dec}})^5/H(t_{\nu\text{-dec}}) \simeq T^{\text{(SM)}}(t_{\nu\text{-dec}}^{\text{(SM)}})^5/H^{\text{(SM)}}(t_{\nu\text{-dec}}^{\text{(SM)}})
\label{eq:nu_decoupling}
\end{equation}
(see also~\cite{Fradette:2017sdd}). 


\subsubsection*{Solution of eq.~(\ref{eq:rhoR_evolutionequation}) prior to neutrino decoupling.}

For $t < t_{\nu\text{-dec}}$, the energy density and pressure in the SM heat bath are given by
\begin{align}
\rho_\text{SM} &= \frac{\pi^2}{30} g_\epsilon^{(T_\nu = T)} (T) T^4 \;\,, \nonumber \\ 
p_\text{SM} &= \frac{\pi^2}{90} g_p^{(T_\nu = T)} (T) T^4 \;\,,
\end{align}
where $g_\epsilon^{(T_\nu = T)} (T)$ and $g_p^{(T_\nu \dfrac{•}{•}= T)} (T)$ are the effective SM degrees of freedom contributing to the energy density and the pressure, with the neutrino temperature $T_\nu$ set equal to the photon temperature $T$. By inserting these expression into eq.~(\ref{eq:rhoR_evolutionequation}), we obtain
\begin{align}
\dot{T} = \frac{m_\phi n_\phi(t)}{\tau_\phi} \frac{1}{G_1(T) T^3} - 3 H(T) T \frac{G_2(T)}{G_1(T)} \quad \quad \quad (t < t_{\nu\text{-dec}}) \;\,,
\label{eq:diffeq_T_priornudec}
\end{align} 
where we have defined
\begin{align}
G_1(T) &\equiv \frac{\pi^2}{30} \left( T \frac{\text{d} g_\epsilon^{(T_\nu = T)} (T)}{\text{d}T} + 4 g_\epsilon^{(T_\nu = T)} (T) \right) \;\,, \label{eq:G1_def}\\
G_2(T) &\equiv \frac{\pi^2}{30} \left(  g_\epsilon^{(T_\nu = T)} (T) + \frac13  g_p^{(T_\nu = T)} (T)\right)\;\,.  \label{eq:G2_def}
\end{align}
We then numerically solve eq.~(\ref{eq:diffeq_T_priornudec}) in order to obtain $T(t)$ prior to neutrino decoupling, i.e.~at $t_0 < t < t_{\nu\text{-dec}}$.


\subsubsection*{Solution of eq.~(\ref{eq:rhoR_evolutionequation}) after neutrino decoupling.}

For $t > t_{\nu\text{-dec}}$, the neutrinos are decoupled from the heat bath, and their effective temperature  simply scales as $T_\nu \propto R^{-1}$:
\begin{align}
T_\nu(t) \big|_{t > t_{\nu\text{-dec}}} = T(t_{\nu\text{-dec}}) \exp \left( - \int_{t_{\nu\text{-dec}}}^{t} \text{d} \lambda \, H(\lambda) \right)\;\,,
\label{neutrino_temperature}
\end{align}
with $T(t_{\nu\text{-dec}})$ being the photon temperature at the time of neutrino decoupling. The energy density and pressure of the SM particles are then given by
\begin{align}
\rho_\text{SM} &= \frac{\pi^2}{30} \left( g_\epsilon^{\text{(vis)}} (T) T^4 + 6 \cdot \frac78 \cdot T_\nu(t)^4 \right) \;\,, \nonumber \\ 
p_\text{SM} &= \frac{\pi^2}{90} \left( g_p^{\text{(vis)}} (T) T^4 + 6 \cdot \frac78 \cdot T_\nu(t)^4 \right) \;\,,
\end{align}
where now $g_\epsilon^{\text{(vis)}} (T)$ and $ g_p^{\text{(vis)}} (T)$ are the `visible' degrees of freedom at a given temperature $T$, i.e.~taking into account all SM particles except the neutrinos. Inserting again into eq.~(\ref{eq:rhoR_evolutionequation}), we obtain
\begin{align}
\dot{T} = \frac{m_\phi n_\phi(t)}{\tau_\phi} \frac{1}{G_1^\text{(vis)}(T) T^3} - 3 H(T) T \frac{G_2^\text{(vis)}(T)}{G_1^\text{(vis)}(T)} \quad \quad \quad (t > t_{\nu\text{-dec}}) \;\,,
\label{eq:diffeq_T_afternudec}
\end{align} 
where $G_1^\text{(vis)}(T)$ and $G_2^\text{(vis)}(T)$ are defined analogously to eqs.~(\ref{eq:G1_def}) and~(\ref{eq:G2_def}) by replacing $g_\epsilon^{(T_\nu = T)} \to g_\epsilon^{\text{(vis)}}$ and $g_p^{(T_\nu = T)} \to g_p^{\text{(vis)}}$. We then numerically solve this differential equation for $T(t)$ starting from $t_{\nu\text{-dec}}$.

Lastly, as already mentioned in the beginning of this section, we determine the `updated' Hubble rate from $H^2 = 8 \pi G/3 \times (\rho_\text{SM} + \rho_\phi)$, and repeat the calculation steps as described above. We find that in all relevant regions of parameter space, three iterations of this procedure are sufficient for converging to a stable solution.


\subsubsection*{Modified baryon-to-photon ratio $\eta$ and $N_\text{eff}^{\text{(CMB)}}$ due to entropy production}

The production of entropy via the decay of $\phi$ further leads to a non-standard time dependence of the baryon-to-photon ratio $\eta = n_B/n_\gamma$. This is implemented in our formalism via the non-standard time-temperature relationship entering the expression for $n_\gamma$, and is fully taken into account in our calculation of nuclear abundances.

Moreover, the fact that after neutrino decoupling the entropy from the decay of $\phi$ is only transferred to the photon bath in general leads to a neutrino-to-photon temperature ratio smaller than the one in the SM~\cite{Fradette:2017sdd}. This leads to a reduced value of $N_\text{eff}^\text{(CMB)}$ at recombination:
\begin{align}
N_\text{eff}^\text{(CMB)} = 3 \left( \frac{T_\nu(t^\text{rec})}{T(t^\text{rec})} \right)^4 \left( \frac{11}{4} \right)^{4/3} \;\,,
\label{eq:NeffCMB_def}
\end{align}
which can directly be evaluated using $T(t)$ and $T_\nu(t)$ as calculated above. We can then confront this to the $2\sigma$ lower bound $N_\text{eff}^\text{(CMB)} > 2.66$ obtained from the latest Planck+BAO data~\cite{Aghanim:2018eyx}. Notice however that this bound can be circumvented by a simple extension of the particle content, e.g.~by postulating the existence of sterile neutrinos which give a positive contribution to $\Delta N_\text{eff}$.


\section{Abundances of light elements}
\label{sec:abundance_of_light_elements}

In order to compare the predicted light element abundances in presence of the additional particle $\phi$ with the corresponding observed values, we need to track the number densities of nuclei throughout the cosmological evolution.
At about one second after the Big Bang the weak interactions freeze out and the number density of neutrons and protons is determined by the mass difference and neutron lifetime which sets the initial
condition for BBN. However, only at about $t \simeq 180\,s$ the production of light nuclei becomes efficient due to the `deuterium bottleneck'. 
At about $t \simeq 10^4\,$s, standard BBN has terminated and the abundances of light elements are no longer changed by the nuclear fusion and spallation reactions.
Subsequent modifications of the nuclear abundances due to photodisintegration reactions are possible, induced by the decay products of $\phi$, happening at $t \gtrsim 10^4\,$s~\cite{Jedamzik:2006xz}.
Note that due to the different times involved, the usual BBN and the subsequent photodisintegration reactions factorise. It is therefore possible to first calculate the nuclear abundances due to nucleosynthesis and 
then consider the abundance changes due to photodisintegration.


\subsection{Nucleosynthesis}
\label{sec:nuclabundance_calculation}

In order to solve the differential equations underlying the formation of light nuclei, we use the rates and the linearisation algorithm that is implemented in \textsc{AlterBBN~v1.4}~\cite{Arbey:2011nf,Arbey:2018zfh}. To incorporate the effect of the additional dark sector particle $\phi$, we replace most of the built-in functions with those resulting from the formalism described in section \ref{sec:cosmo_evolution}, i.e.~we update
\begin{itemize}
	\item The time-temperature relation and the modified Hubble rate (cf.~eqs.~\eqref{eq:diffeq_T_priornudec} and~\eqref{eq:diffeq_T_afternudec}).
	\item The baryon-to-photon ratio, which we fix to the value $\eta = 6.1\times 10^{-10}$~\cite{Ade:2015xua} at the time of recombination $t_\text{CMB} \simeq 10^{12}\;\mathrm{s}$. Before recombination (or more precisely before the decay of the mediator), however, this condition implies $\eta \geq 6.1\times 10^{-10}$, since the decay of the mediator produces additional entropy and therefore leads to a decrease of the baryon-to-photon ratio.
	\item The neutrino temperature, including a neutrino decoupling time that is different from the Standard Model BBN scenario due to the modified Hubble rate (cf. eq.~\eqref{eq:nu_decoupling}).
\end{itemize}
Depending on the lifetime and the mass of the particle $\phi$, the resulting abundances may still subsequently be altered by photodisintegration as discussed below.


\subsection{Late-time modifications of nuclear abundances via photodisintegration}
\label{sec:photodis}

\subsection*{Calculation of the non-thermal photon spectrum}

Given that the decay products of $\phi$ are photons and charged leptons, they will induce an electromagnetic cascade once injected. It has been known for a long time~\cite{Cyburt:2002uv} that for sufficiently large injection energies those cascades lead to a quasi-universal photon spectrum depending only on the injection time and total energy injected,
\begin{equation*}
f_\gamma^{\text{(uni.)}}(E)\sim
\begin{cases}
K_0 \left( \frac{E}{E_X} \right)^{-3/2} &\text{for $E < E_X$}\\
K_0 \left( \frac{E}{E_X} \right)^{-2} &\text{for $E_X \leq E \leq E_C$}\\
0 &\text{for $E > E_C$}
\end{cases}
\label{eq:fgamma_univ}
\end{equation*}
with $K_0 = E_0E_X^{-2}[2 + \ln(E_C/E_X)]^{-1}$, $E_C \simeq m_e^2/(22T)$ and $E_X \simeq m_e^2/(80T)$~\cite{Kawasaki:1994sc}.
However, it has been pointed out that this approximation breaks down if the energy of the initial decay products is below the effective cutoff energy for the production of electron-positron pairs $E_C$, i.e.~if the decaying particle is too light~\cite{Poulin:2015opa}.
This is the case for a sizeable part of the parameter space we are interested in, which makes it mandatory
to calculate the non-thermal spectrum from scratch by solving the corresponding cascade equations, thereby taking into account all the processes that alter the non-thermal spectra of the injected particles.
In the following, we thus study the coupled evolution equations of all particles that can emerge in electromagnetic decays, i.e.~photons, electrons and positrons, generalising the discussion in~\cite{Poulin:2015opa} where only the photon spectrum was considered.
Setting $\partial f_X(E)/\partial t = 0$ and using $\Gamma_X \gg H$~\cite{Kawasaki:1994sc,Jedamzik:2006xz}, the relevant integral equations \textcolor{black}{for the number $f_X$ of particles $X \in \left\{ \gamma, e^-, e^+ \right\}$ per unit volume and energy} can be written as (suppressing the $T$ dependence of all quantities)
\begin{align}
f_X(E) = \frac{1}{\Gamma_X(E)} \left( S_X(E) + \int_{E}^{\infty} \text{d} E' \sum_{X'} \left[ K_{X' \to X} (E, E') f_{X'}(E') \right] \right) \,.
\label{eq:fX_recursive}
\end{align}
Here, $\Gamma_X(E)$ is the total interaction rate of particle $X$ at energy $E$, $K_{X' \to X} (E, E')$ is the differential interaction rate for scattering/conversion of particle $X'$ with energy $E'$ to particle $X$ with energy $E$ and $S_X(E)$ is the source term for the production of particle $X$ with energy $E$. The latter is given by
\begin{align}
S_X(E) = S_X^{(0)} \delta(E-E_0) + S_X^\text{(FSR)}(E) \,,
\label{eq:SXE_definition}
\end{align}
with the first term corresponding to the monochromatic energy injection of the particle $X$ with energy $E_0 = m_\phi/2$\footnote{This implies that the particle $\phi$ has become non-relativistic when it decays. Given that photodisintegration can only happen for $m_\phi \gtrsim 4\,$MeV and at times $t_\phi \gtrsim 10^4\,$s, this is always the case and does not correspond to an additional assumption.}:
\begin{equation}
S_\gamma^{(0)} = \text{BR}_{\gamma\gamma} \times \frac{2n_\phi}{\tau_\phi}\;\quad \text{and} \quad S_{e^-}^{(0)} = S_{e^+}^{(0)} = \text{BR}_{e^+ e^-} \times \frac{n_\phi}{\tau_\phi}\;\,,
\end{equation}
where $\text{BR}_{e^+ e^-}$ and $\text{BR}_{\gamma\gamma} = 1-\text{BR}_{e^+ e^-}$ are the branching ratios into electron-positron pairs and photons respectively. In addition, following~\cite{Forestell:2018txr}, the second term in eq.~(\ref{eq:SXE_definition}) takes into account final state radiation (FSR) of photons for the case of $\phi$ decaying into $e^+ e^-$\cite{Mardon:2009rc, Birkedal:2005ep}:
\begin{equation}
S_\gamma^\text{(FSR)}(E) = \text{BR}_{e^+e^-} \times \frac{n_\phi}{\tau_\phi E_0}\times \frac{\alpha}{\pi} \frac{1+(1-x)^2}{x}\ln\left( \frac{4E_0^2(1-x)}{m_e^2} \right) \times \Theta\left( 1 - \frac{m_e^2}{4E_0^2} - x \right)
\label{eq:SFSR}
\end{equation}
with $x=E/E_0$.

For the solution of eq.~(\ref{eq:fX_recursive}), it is beneficial to split off the delta function from the rest of the spectrum to achieve stable numerical calculations and analytically reinsert it at a later stage. We therefore define 
\begin{align}
F_X(E) \equiv f_X(E) - \frac{S_X^{(0)} \delta(E-E_0)}{\Gamma_X(E)}
\label{F_f_relation}
\end{align}
Inserting this definition into eq.~(\ref{eq:fX_recursive}) we find
\begin{align}
\Gamma_X(E) F_X(E) = S_{X}^\text{(FSR)}(E) + \sum_{X'} \left[ \frac{K_{X' \to X} (E, E_0) S_{X'}^{(0)}}{\Gamma_{X'} (E_0)} + \int_{E}^{\infty} \text{d} E' K_{X' \to X} (E, E') F_{X'}(E')\right] \;\,.
\label{eq:FX_recursive}
\end{align}
This is a coupled Volterra integral equation of type 2 for the three different spectra $F_X$, which can be solved numerically by using a discretisation method similar to the one used in~\cite{Jedamzik:2006xz}.

For the rates $\Gamma_X$ and the kernels $K_{X'\rightarrow X}$, the dominant scattering processes on the thermal photons $\gamma_\text{th}$ as well as on the background electrons $e^-_\text{th}$ and nuclei $N$ are:
\begin{enumerate}
	\item Double photon pair creation $\gamma + \gamma_\text{th} \rightarrow e^+ + e^-$
	\item Photon-photon scattering $\gamma + \gamma_\text{th} \rightarrow \gamma + \gamma$
	\item Bethe-Heitler pair creation $\gamma + N \rightarrow e^+ + e^- + N$ with $N \in \{{}^1\text{H}, {}^4\text{He} \}$
	\item Compton scattering $\gamma + e^-_\text{th} \rightarrow \gamma + e^-$\\
	Since there are (almost) no background positrons present at the time of photodisintegration, it is this reaction which is actually responsible for the difference of the electron and the positron spectrum.
	\item Inverse Compton scattering $e^{\pm} + \gamma_\text{th} \rightarrow e^{\pm} + \gamma$
\end{enumerate}
\begin{figure}
	\begin{center}		
	        \hspace*{-0.7cm}
		\includegraphics[scale=0.56]{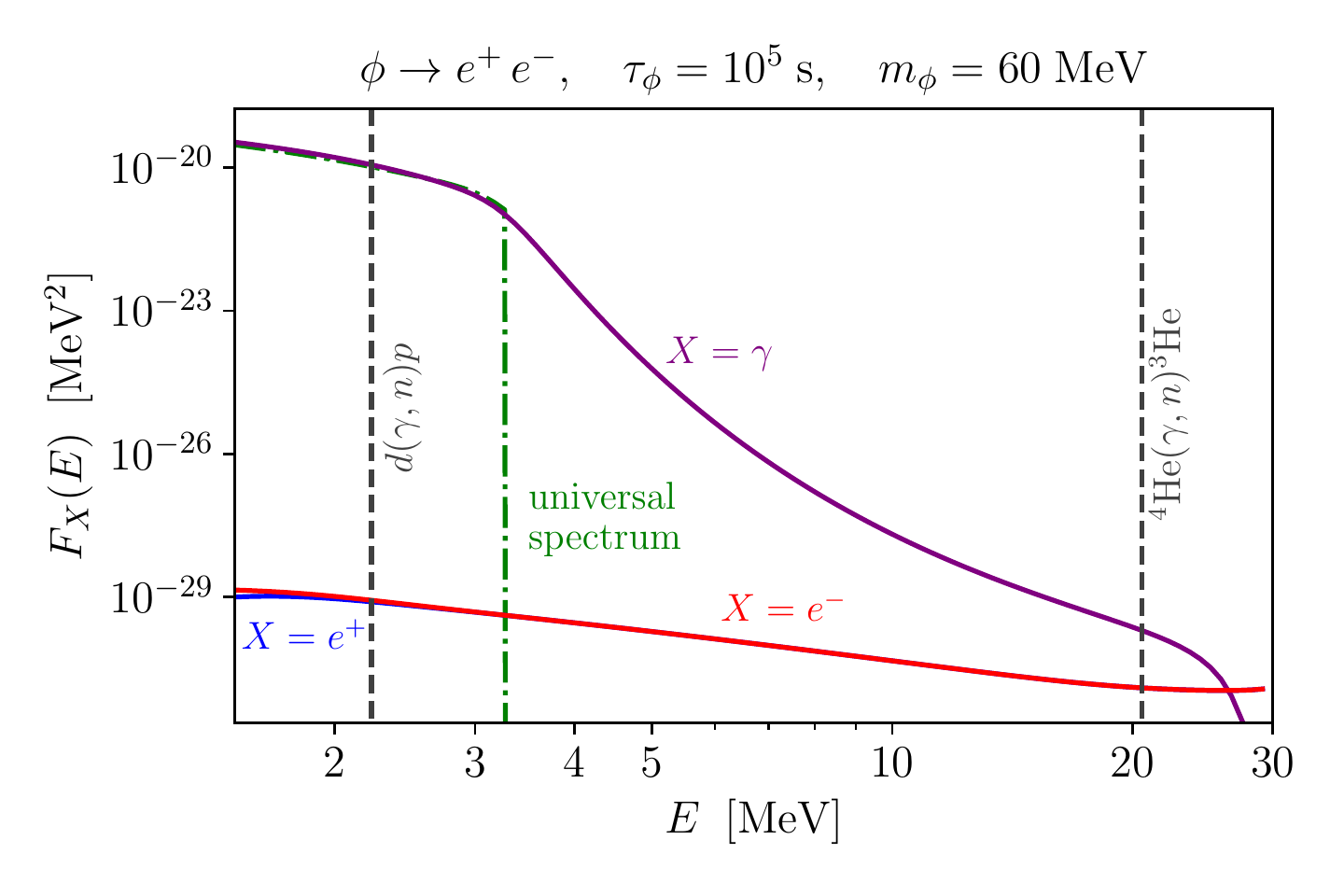}\hspace*{-0.2cm}
		\includegraphics[scale=0.56]{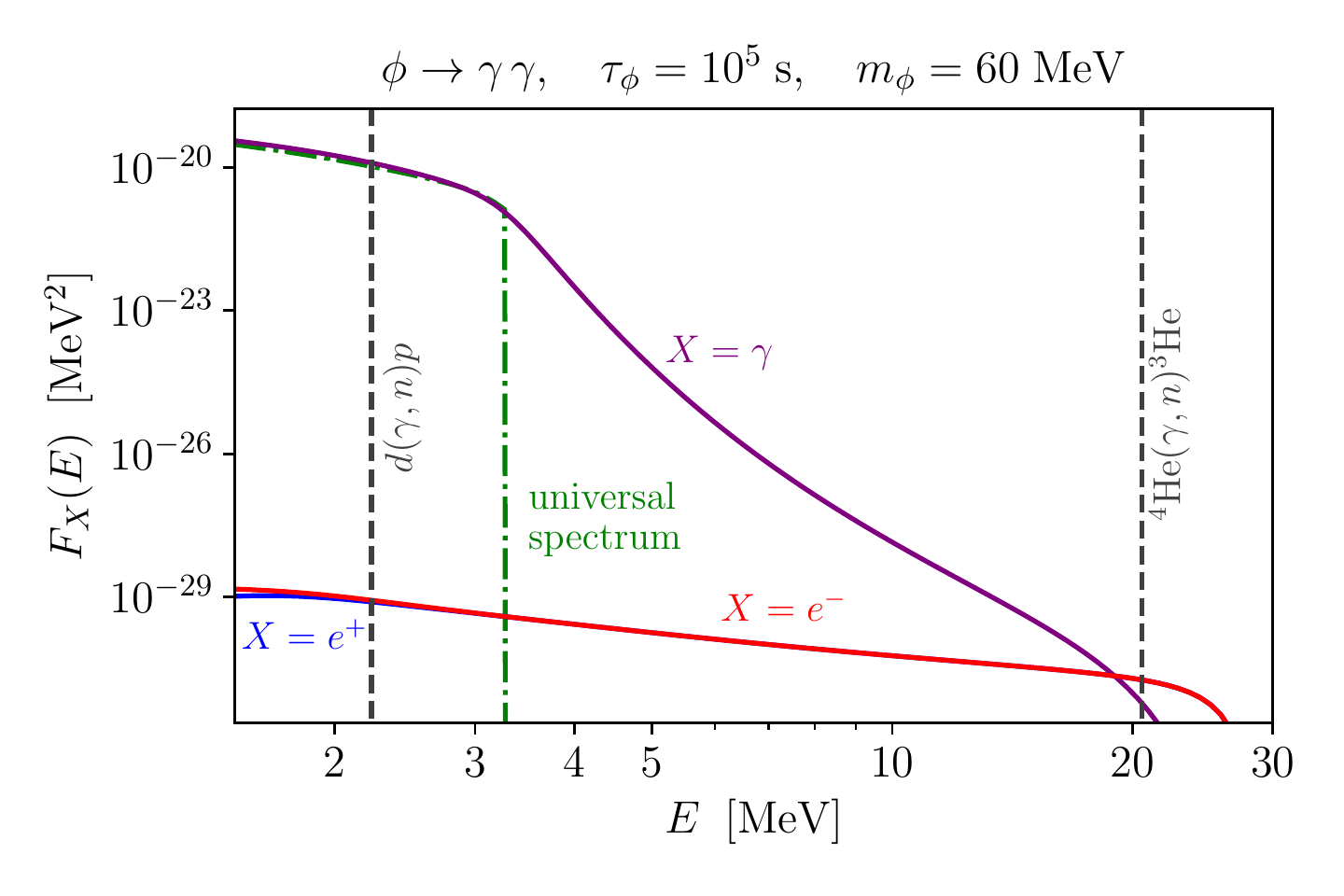}
		\hspace*{-0.7cm}
		\includegraphics[scale=0.56]{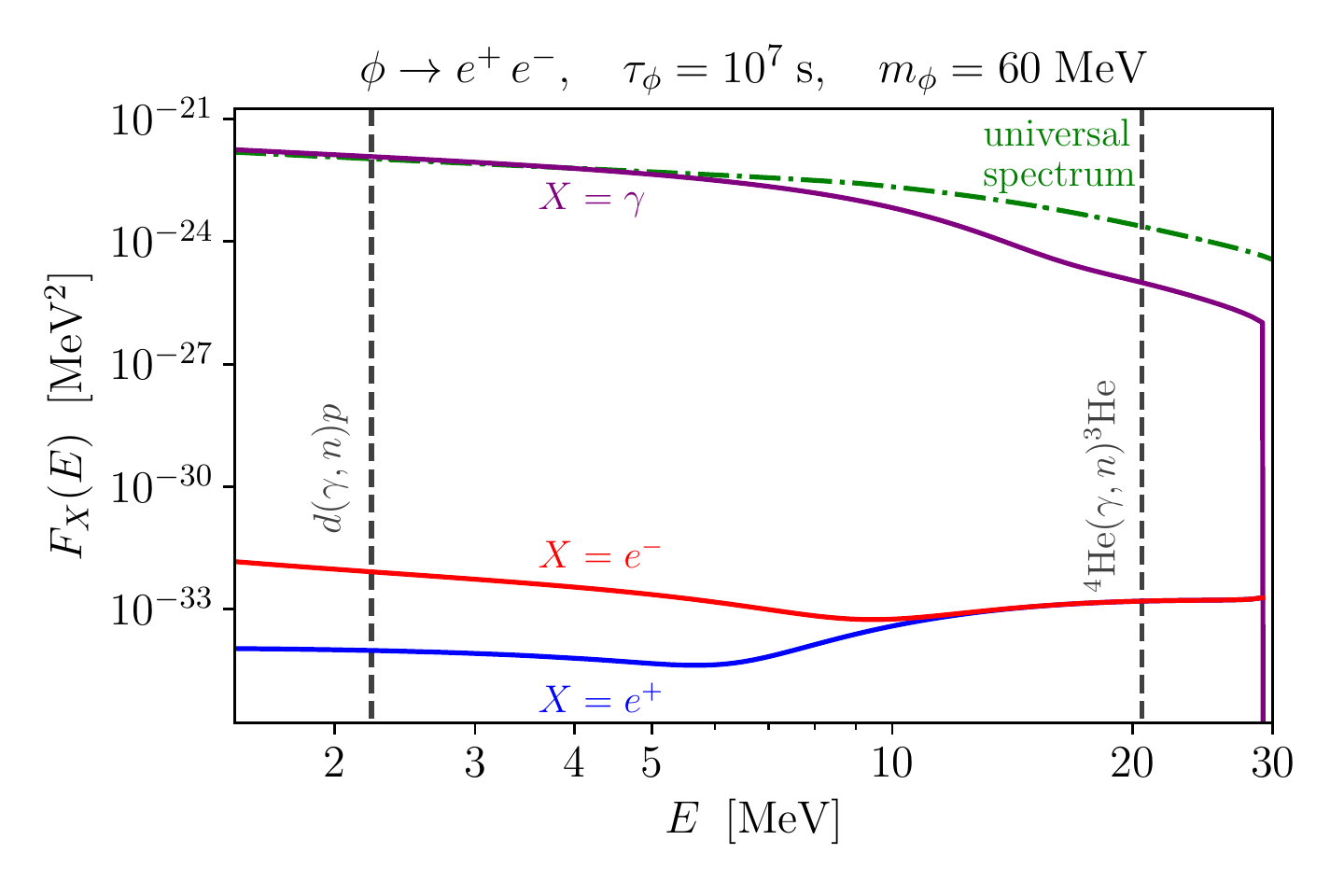}\hspace*{-0.2cm}
		\includegraphics[scale=0.56]{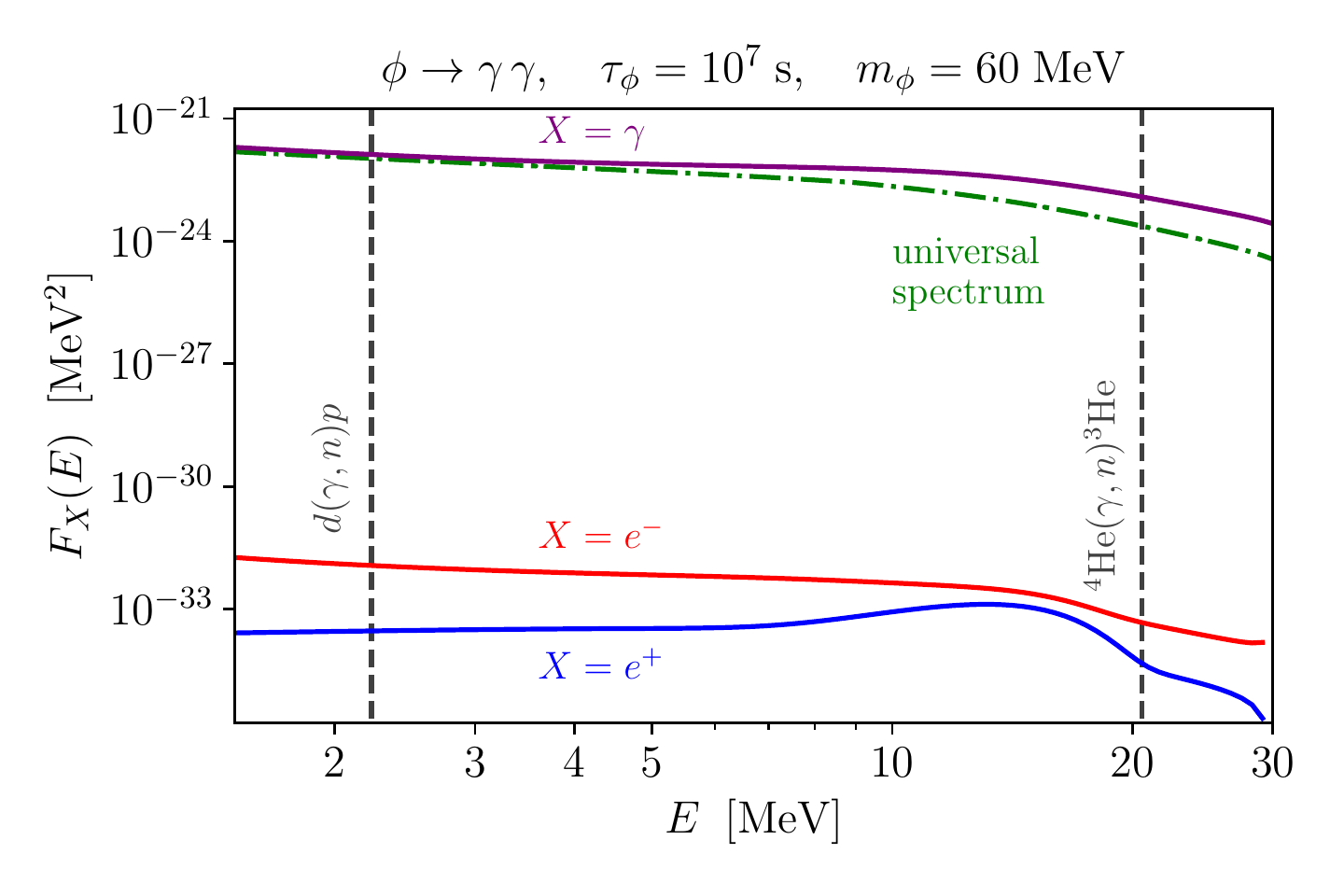}\\
	\end{center}
	\caption{Comparison of the differential spectra for $\phi \rightarrow e^+ e^-$ (left) and $\phi \rightarrow \gamma \gamma$ (right) for two different lifetimes, $\tau_\phi=10^5$s (top) and $\tau_\phi=10^7$s (bottom) with $m_\phi = 60\,\mathrm{MeV}$ and $\textcolor{black}{n_\phi/n_\gamma|_{T = T^\text{cd} = 10\,\mathrm{GeV}} = 10^{-5}}$, evaluated at a temperature $T$ corresponding to $t \simeq \tau_\phi$.
	For reference also the universal spectrum as well as the photodisintegration thresholds of deuterium and $^4$He are shown.}
	\label{fig:spectra_comparison}
\end{figure}
Given that a number of typos are present in the literature, we provide the rates and kernels for all of these processes in appendix \ref{app:rates_cascade} for convenience. Using the full set of relevant rates, we calculate the non-thermal spectra for all relevant temperatures. As an example, in figure~\ref{fig:spectra_comparison} we present the photon, electron and positron spectra $F_X(E)$ for the case of a particle with mass $m_\phi = 60\,$MeV and lifetime $\tau_\phi=10^5$s (top panels) and $\tau_\phi=10^7$s (bottom panels), decaying into $e^+ e^-$ (left panels) and $\gamma \gamma$ (right panels). The abundance of $\phi$ at the reference temperature $T = 10\,$GeV is fixed to $n_\phi = 10^{-7} n_\gamma$. Also note that the spectra change with temperature, and for definiteness are shown for $T = 3.6 \times 10^{-3}$\,MeV ($3.6 \times 10^{-4}$\,MeV) in the upper (lower) panels, corresponding to $t \simeq \tau_\phi$.

In all cases considered in figure~\ref{fig:spectra_comparison}, we find that the electron and positron spectra are strongly suppressed with respect to the photon spectrum, even for the case where $\phi$ decays into electron-positron pairs. This can be understood by noting that (below the double photon pair creation threshold) the production of high-energetic $e^\pm$ can only proceed via Compton scattering off background electrons or via Bethe-Heitler scattering off background nuclei, both of which are strongly suppressed due to the low density of targets. On the other hand, high-energetic photons are frequently produced from inverse Compton scattering on the much more abundant background photons. For the case of $\tau_\phi = 10^5\,$s shown in the upper panels of figure~\ref{fig:spectra_comparison}, our full calculation of the cascade process reproduces remarkably well the universal spectrum of photons below the cutoff energy $E_C$, which provides an important consistency check of our approach. \textcolor{black}{Moreover, for energies $E > E_C$ the universal spectrum vanishes by construction, while our calculation includes the exponentially suppressed spectrum of photons which do not fully convert their energy to values below $E_C$ via the production of electron-positron pairs.}

On the other hand, for the parameters shown in the lower panels the condition $T \gtrsim m_e^2/(22E_0)$ is no longer fulfilled, and consequently our photon spectra deviate substantially from the universal spectrum. \textcolor{black}{More precisely, for $E \gtrsim 5\,$MeV the universal spectrum overestimates the photon flux for the case of decays into $e^+ e^-$ (lower left panel of figure~\ref{fig:spectra_comparison}), while it underestimates it for decays into $\gamma \gamma$ (lower right panel). Qualitatively, this can be understood as follows: 
in the derivation of the universal spectrum, one assumes that the energy injected by high-energy photons (electrons) is efficiently transferred to the other species according to eq.~\eqref{eq:fgamma_univ}. 
However, when the underlying assumption $T \gtrsim m_e^2/(22E_0)$ is not satisfied, this process is less efficient, and for the case of decays into $e^+ e^-$ a smaller fraction of the energy is transferred to the photons, resulting in the suppressed photon spectrum visible in the lower left panel of figure~\ref{fig:spectra_comparison}. Conversely, for decays into $\gamma \gamma$ the same argument implies a larger flux of photons compared to the prediction of the universal spectrum.}


\subsection*{Evolution of the light-element abundances during photodisintegration}
The late-time modification of the nuclear abundances caused by the process of photodisintegration is described by the following differential equation~\cite{Cyburt:2002uv, Poulin:2015opa} (again dropping the $T$ dependence of all quantities):
\begin{align}
\left( \frac{\text{d} T}{\text{d} t} \right) \frac{\text{d} Y_X}{\text{d} T} & = \sum_{N_i} Y_{N_i} \int_{0}^{\infty}\text{d} E\; f_\gamma(E)\sigma_{\gamma + N_i \rightarrow X}(E) \nonumber \\
&\;\, - Y_X \sum_{N_f} \int_{0}^{\infty}\text{d} E\; f_\gamma(E)\sigma_{\gamma + X \rightarrow N_f}(E)
\end{align}
with $Y_X = n_X/n_b$ and $X \in \{ p, n, {}^2\text{H}, {}^3\text{H}, {}^3\text{He}, {}^4\text{He}, \dots \}$. Substituting as in eq.~\eqref{F_f_relation}, we find
\begin{align}
\left( \frac{\text{d} T}{\text{d} t} \right) \frac{\text{d} Y_X}{\text{d} T} & = \sum_{N_i} Y_{N_i} \left[ \int_{0}^{\infty}\text{d} E\; F_\gamma(E)\sigma_{\gamma + N_i \rightarrow X}(E) + \frac{\sigma_{\gamma + N_i \rightarrow X}(E_0)S_\gamma^{(0)}}{\Gamma_\gamma(E_0)}\right] \nonumber \\
& \;\, - Y_X \sum_{N_f} \left[ \int_{0}^{\infty}\text{d} E\; F_\gamma(E)\sigma_{\gamma + X \rightarrow N_f}(E) + \frac{\sigma_{\gamma + X \rightarrow N_f}(E_0)S_\gamma^{(0)}}{\Gamma_\gamma(E_0)} \right]
\label{eq:pdi}
\end{align}
with the time-temperature relation $\text{d}T/\text{d}t$ from eqs.~\eqref{eq:diffeq_T_priornudec} and~\eqref{eq:diffeq_T_afternudec}.
This is a linear, ordinary coupled differential equation for $Y_X$, which can be solved analytically (expect for the integrals over the rates). As discussed above, the initial conditions are taken to be the abundances at the end of standard nucleosynthesis, i.e.~those calculated by our modified version of \textsc{AlterBBN}. In all cases of interest we can neglect reactions that involve elements heavier than $^4$He, as their effect is negligible in this context. We implement the rates for the reactions \textit{1-9} from~\cite{Cyburt:2002uv}, but modify the prefactor of reaction \textit{7} from 17.1mb to 20.7mb as suggested by~\cite{Jedamzik:2006xz}, in order to match the most recent \textsc{EXFOR} data\footnote{\url{https://www-nds.iaea.org/exfor/exfor.htm}.}. Note that the electron/positron spectra influence the nuclear abundances only indirectly via their effect on the photon spectrum.
\begin{figure}
	\begin{center}
		\hspace*{-0.6cm}
		\includegraphics[scale=0.7]{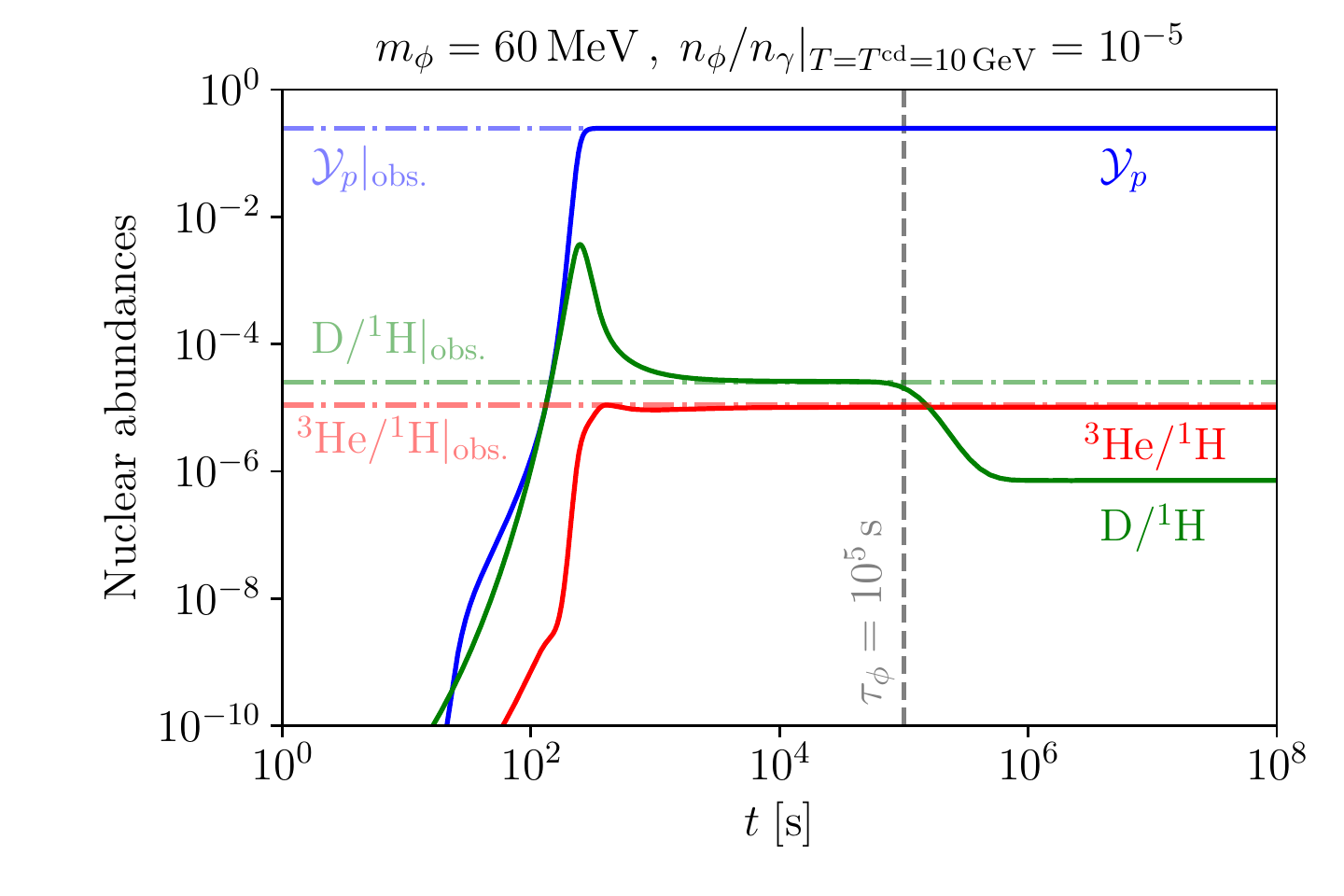}
	\end{center}
	\caption{Evolution of the nuclear abundances during nucleosynthesis and photodisintegration, assuming $m_\phi = 60\,$MeV, $n_\phi/n_\gamma \big|_{T = T^\text{cd} = 10\,\text{GeV}} = 10^{-5}$, $\tau_\phi = 10^5\,$s and decay into electron-positron pairs.}
	\label{fig:bbn_pdi_evolution}
\end{figure}

In figure~\ref{fig:bbn_pdi_evolution} we show for illustration the time evolution of the abundances of D, ${}^3$He and ${}^4$He for a particle with mass $m_\phi = 60\,$MeV  and an initial abundance $n_\phi = 10^{-5} n_\gamma$ at a reference temperature $T = 10\,$GeV, decaying into $e^\pm$ with a lifetime $\tau_\phi = 10^5\,$s. As explained in the beginning of this section, the nuclear abundances are strongly suppressed before $t \simeq 180\,$s due to the deuterium bottleneck, and reach a plateau before $t \simeq 10^4\,$s, corresponding to the end of standard nucleosynthesis. After the decay of $\phi$ at $t \gtrsim 10^5\,$s, photodisintegration then leads to a decrease of the deuterium abundance according to eq.~(\ref{eq:pdi}).


\subsection{Comparison to observations}
To evaluate whether a given point in parameter space is viable, we compare
the predicted present day abundances $Y_X(T \rightarrow 0)$ to the most recent compilation of observations~\cite{Olive:2016xmw, Bania:2002yj}:
\begin{align}
& \mathcal{Y}_\text{p} \quad & (2.45 \pm 0.04) \times 10^{-1}  \; \,,\\
& \text{D}/{}^1\text{H} \quad & (2.53 \pm 0.04) \times 10^{-5} \;\,, \\
& {}^3\text{He}/{}^1\text{H} \quad & (1.1 \pm 0.2) \times 10^{-5} \label{eq:3HeH_abundance} \;\,.
\end{align}
For the constraints after photodisintegration, the bound on ${}^3\text{He}/^1\text{H}$ turns out to be of particular importance, which is why we add this measurement to the set of observations used in~\cite{Hufnagel:2017dgo}. \textcolor{black}{However, in contrast to D and $^4$He, the abundance of $^3$He is solely inferred from high-metallicity environments, making the connection with the primordial abundance less clear~\cite{VangioniFlam:2002sa,Olive:2016xmw}. Following~\cite{Bania:2002yj}, in this work we assume that stellar processes can only lead to additional production of $^3$He, and hence employ an upper bound on the primordial abundance of $^3\text{He}/{}^1\text{H}$ given by eq.~(\ref{eq:3HeH_abundance}). Note however that in~\cite{Kawasaki:2004qu} it has been argued that  the ratio $^3\text{He}/\text{D}$ is a more robust probe for the primordial abundance of $^3$He. We find that this would weaken the corresponding constraints on the abundance of $\phi$ by a factor of a few; however, as argued below, the bound from $^3\text{He}$ is in any case only relevant for very small abundances of $\phi$, and in particular does not impact any of our conclusions for thermal abundances.}
Furthermore, as in~\cite{Hufnagel:2017dgo} we conservatively do not apply any bound on the lithium abundance given the well known discrepancy with the standard BBN prediction~\cite{Fields:2011zzb} as well as the corresponding large systematic uncertainties~\cite{Korn:2006tv}.

To take into account the theoretical uncertainties on the nuclear rates, we utilise the $\pm 1 \sigma$ high and low values of the nuclear reaction rates that are implemented in {\tt AlterBBN} as described in~\cite{Hufnagel:2017dgo}. Specifically, we compute three different values for the abundances after BBN, $Y_X^{\text{(BBN)}}$, $Y_{X, +1\sigma}^{\text{(BBN)}}$ and $Y_{X, -1\sigma}^{\text{(BBN)}}$ and solve eq.~\eqref{eq:pdi} for these three different initial conditions. We denote the corresponding abundances after photodisintegration by $Y_X^{\text{(PDI)}}$, $Y_{X, +1\sigma}^{\text{(PDI)}}$ and $Y_{X, -1\sigma}^{\text{(PDI)}}$ and define the observable abundance ratios as
\begin{align}
& R_X^\text{(PDI)} \equiv X/{}^1\text{H} =  Y_X^\text{(PDI)}/Y_{{}^1\text{H}}^\text{(PDI)}\quad\text{for}\quad X \in \{\text{D}, {}^3\text{He}\} \,,\\
\text{and}\qquad & R_{{}^4\text{He}}^\text{(PDI)} \equiv \mathcal{Y}_p = 4\cdot Y_{{}^4\text{He}}^\text{(PDI)}\;\,.
\end{align} 
We approximate the theoretical $1\sigma$ error on each abundance ratio via
\begin{equation}
\sigma_{R_X}^{\text{th}} = \min_i\left(\left|R_{X}^{\text{(PDI)}} - R_{X, +\sigma}^{\text{(PDI)}}\right|, \left|R_{X}^{\text{(PDI)}}- R_{X, -\sigma}^{\text{(PDI)}}\right|\right) 
\end{equation}
and consider a given parameter point to be excluded at the $2\sigma$ level if
\begin{equation}
\Delta_{R_X} \equiv \left| R_X - R_X^{\text{obs}} \right| \bigg/ \sqrt{ \left( \sigma_{R_X}^{\text{th}} \right)^2 + \left( \sigma_{R_X}^{\text{obs}} \right)^2 } \geq 2
\label{eq:Delta_Ri}
\end{equation}
for at least one abundance ratio $R_X$.
We fix the neutron lifetime to its best fit value, $\tau_n = 880\;\mathrm{s}$~\cite{2016RvMP...88a5004C}, having checked that a variation of $\tau_n$ within its uncertainties does not lead to a significant change of the abundance ratios.

\section{Results}
\label{sec:results}

\subsection{Upper limits on the abundance of $\phi$}
\label{sec:upperlimits_abundance}

In the following, we present the $2\sigma$ upper bounds from the combination of nucleosynthesis and photodisintegration on the abundance of $\phi$, together with the CMB constraint on $N_\text{eff}^{\text{(CMB)}}$. To this end, let us first note that in the discussion of the cosmological evolution of $\phi$ and its decay products in section~\ref{sec:cosmo_evolution} we did not further specify the initial phase space distribution $f_\phi(t_0, E) = f_\phi^{(0)}(E)$. If the particle decays while being non-relativistic, BBN and CMB observables are only sensitive to the number density $n_\phi$ making such a specification unnecessary. However, if the particle decays while being semi- or ultra-relativistic, the total energy density as well as the relation between the actual and proper decay time explicitly depend on $f_\phi^{(0)}(E)$. For definiteness, in the following we will thus assume that $\phi$ chemically decoupled at $t_0 \equiv t^\text{cd}$ (corresponding to a photon temperature $T^\text{cd}$), with the distribution function of $\phi$ being proportional to a thermal distribution with temperature $T_\phi^\text{cd}$. The overall normalisation $\propto n_\phi$ is left as a free parameter to be constrained by data. As a benchmark choice we employ the case of $T_\phi^\text{cd}/T^\text{cd} = 1$, but we also show results for other temperature ratios. Studying non-thermal initial distribution functions for the decaying particle is possible using the formalism developed in section~\ref{sec:cosmo_evolution}, but is beyond the scope of this work. Also, for now fix the chemical decoupling temperature of $\phi$ to be $T^\text{cd} = 10\,$GeV, 
but we later comment on how to (trivially) rescale our results to different values of the decoupling temperature. Furthermore we assume the mediator to decay exclusively into $e^+ e^-$ in this section; given that photodisintegration is anyway only possible for $m_\phi \gtrsim 4\,$MeV this is a very natural assumption from a model-independent point of view as it would automatically arise from Higgs mixing. Note however that the bounds arising from nucleosynthesis itself, i.e.~from the modified Hubble rate and/or entropy production are insensitive to whether $\phi$ decays into photons or electron-positron pairs.
In appendix~\ref{app:plot_collection} we provide additional results for $\phi$ decaying into photons as well as for a large number of different masses. 

\begin{figure}
\begin{center}
\hspace*{-0.7cm}
\includegraphics[scale=0.68]{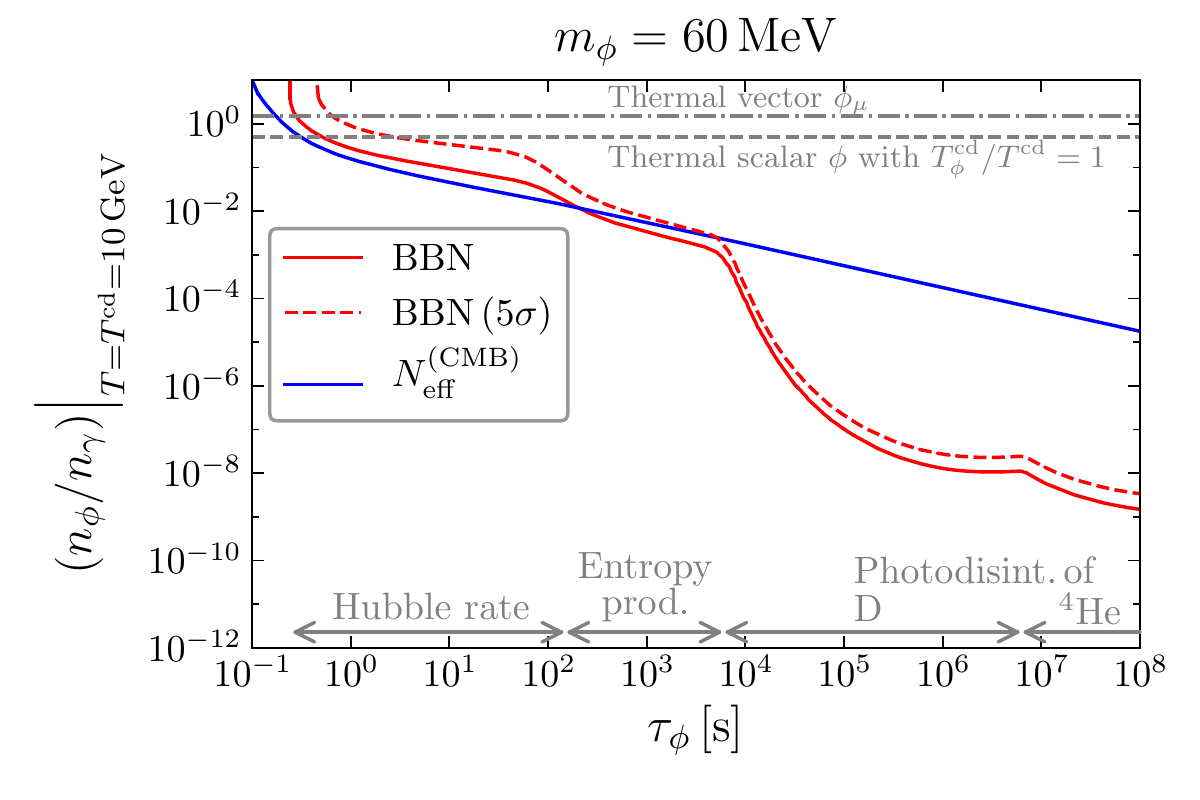}\hspace*{-0.2cm}
\includegraphics[scale=0.68]{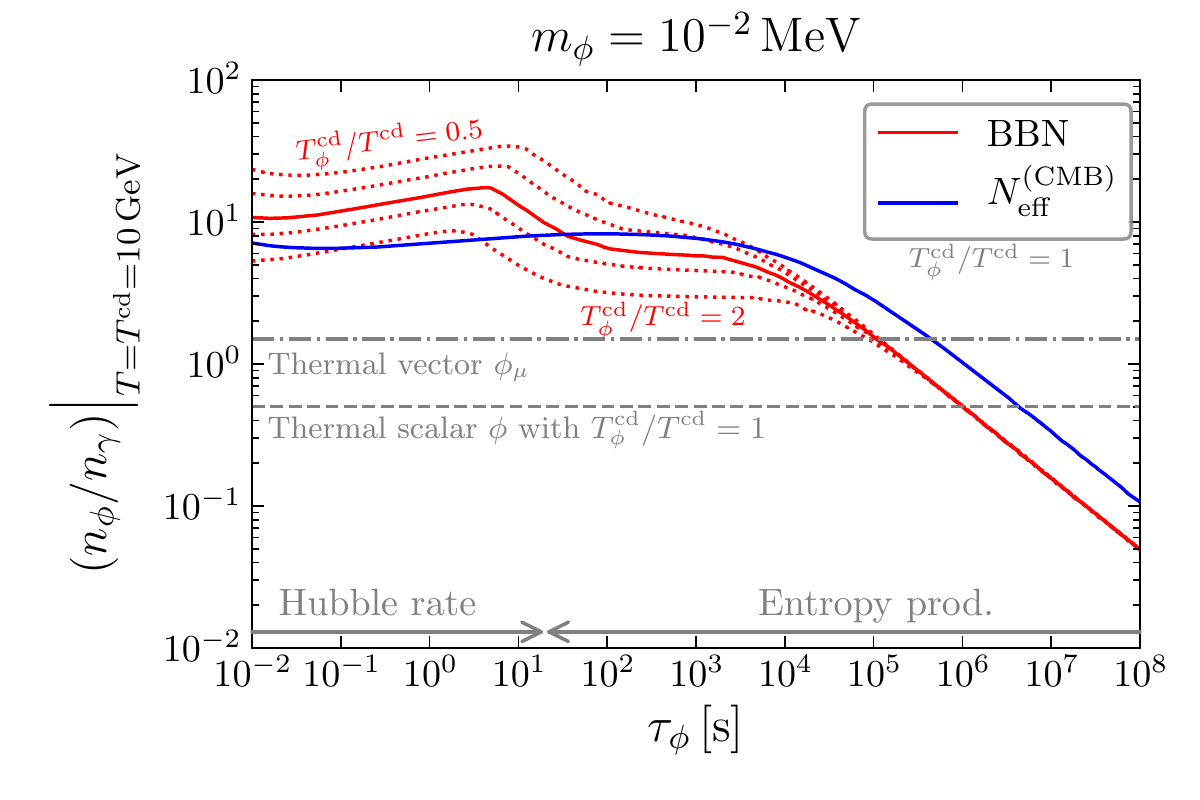}
\end{center}
\caption{BBN and $N_\text{eff}^\text{(CMB)}$ limits \textcolor{black}{(at $2 \sigma$)} on the abundance of $\phi$ for $m_\phi=60$ MeV (left) and $m_\phi=0.01$MeV (right) 
as a function of the lifetime $\tau_\phi$. \textcolor{black}{For comparison, in the left panel the red dashed curve shows the $5\sigma$ bound from BBN.} The thermal abundance is given by 3/2 (1/2) times the photon abundance for a massive vector (scalar) as indicated by the grey dashed lines. Depending on the lifetime $\tau_\phi$, the limit is dominated by different effects, as indicated by the grey arrows at the bottom of the panels. Note the different scalings of the vertical axis for the two different masses.}
\label{fig:results_fixed_mphi}
\end{figure}
In figure~\ref{fig:results_fixed_mphi} we show the relevant exclusion limits depending on the abundance measure  $(n_\phi/n_\gamma)|_{T=T^\text{cd}=10\mathrm{GeV}}$ and the lifetime $\tau_\phi$ of the mediator for two different values of the mediator mass $m_\phi =60\;\mathrm{MeV}$ (left) and $m_\phi =0.01\;\mathrm{MeV}$ (right). The thermal abundance is given by the photon abundance times the ratio of the relevant degrees of freedom, i.e.\ 3/2 (1/2) for a massive vector (scalar) as indicated by the grey dashed lines. In general the temperatures in the dark and visible sector need not be the same and for a temperature ratio $T_\phi^\text{cd}/T^\text{cd} \ne 1$ the thermal abundance scales with $(T_\phi^\text{cd}/T^\text{cd})^3$.

For a mediator mass of $60\,$MeV, the BBN bounds shown in red are sensitive to three different effects, each of them becoming dominant for a different range of lifetimes, as indicated by the grey arrows at the bottom of figure~\ref{fig:results_fixed_mphi}.
For {\it small lifetimes}, $\tau_\phi \lesssim 200\,$s, the limit dominantly arises from the increased Hubble rate (or equivalently from the modified time-temperature relationship) after neutron-proton freeze-out at $t \simeq 1\,$s, induced by the extra energy density associated to $\phi$. It can also be seen that the limit on the abundance of $\phi$ becomes stronger with increasing lifetime. This is because for larger values of $\tau_\phi$, the particle has more time to profit from the scaling $\rho_\phi \propto R^{-3}$ of non-relativistic matter compared to the one of radiation $\rho_\text{SM} \propto R^{-4}$; hence, the energy density at the time of decay increases with its lifetime. 
For {\it intermediate lifetimes}, $200 \, \text{s} \lesssim \tau_\phi \lesssim 10^4\,$s, the limit dominantly arises from the additional entropy that is produced during the decay of $\phi$. More precisely, in this regime, the produced electron-positron pairs quickly thermalise with the SM heat bath, thus decreasing the value of the baryon-to-photon ratio during BBN. However, since the value of $\eta$ at the time of recombination is known, $\eta_\text{CMB} \simeq 6.1 \times 10^{-10}$, the corresponding value during BBN must have been larger, which may result in nuclear abundances that are in conflict with the respective observational values. Numerically, we find that the baryon-to-photon ratio during BBN must not exceed the CMB value by more than about 5\%, $\eta \lesssim 1.05 \, \eta_\text{CMB}$. 
As before, the energy density and therefore the limit on the abundance of $\phi$ becomes stronger with increasing lifetime. For $\tau_\phi \lesssim 200\,$s, the production of entropy and thus the decrease of the baryon-to-photon ratio occurs prior to the time when light nuclei are dominantly produced (which happens only at $t \gtrsim 180\,$s due to the `deuterium bottleneck'). In this case $\eta \simeq \eta_\text{CMB}$ during the time most relevant for BBN and consequently the bound from $\eta_{\text{BBN}} \neq \eta_\text{CMB}$ (entropy production) vanishes.
For {\it large lifetimes},  $\tau_\phi \gtrsim 10^4\,$s, the dominant bound comes from photodisintegration. Specifically, for $10^4\,\text{s} \lesssim \tau_\phi \lesssim 8 \times 10^6\,$s (and the mass considered) the most relevant process is the destruction of deuterium, quickly leading to a tension with the $2\sigma$ lower bound on the observationally inferred deuterium abundance D/$^1$H. For $\tau_\phi \gtrsim 8 \times 10^6\,$s, photodissociation of $^4$He becomes efficient, mainly due to the photons in the tail of the FSR spectrum at $E_\gamma \lesssim m_\phi/2$ (see eq.~(\ref{eq:SFSR})). Finally, it can be seen from the left panel of figure~\ref{fig:results_fixed_mphi} that for $\tau_\phi \lesssim 10^4\,$s the lower bound on $N_\text{eff}^\text{(CMB)}$ (c.f.~eq.~(\ref{eq:NeffCMB_def})) gives a constraint comparable to the one from BBN, while it is much less constraining for larger lifetimes. As already mentioned in section~\ref{sec:cosmo_evolution}, it is also important to keep in mind that this bound can be circumvented by extending the particle content of the model e.g.~with sterile neutrinos.

For the case of a much smaller mediator mass, $m_\phi=0.01\,$MeV (shown in the right panel of figure~\ref{fig:appendix_plots}), the injected energy is below the binding energy of all relevant nuclei, which is why photodisintegration is irrelevant for such a scenario. Consequently, depending on the range of lifetimes, the limit either arises from the increased Hubble rate or from the production of additional entropy. Specifically the entropy bound starts to dominate already at lifetimes of $\tau_\phi \simeq 20\,s$. This is because for $m_\phi=0.01 \,$MeV, $\phi$ decays while being relativistic, and thus the actual time of decay $t^\text{decay}$ is larger than the proper lifetime by a Lorentz boost of order $T(t^\text{decay})/m_\phi \simeq 10$. Notice that this effect is fully taken into account in our analysis, as we do not make any approximation of $\phi$ being ultra- or non-relativistic during its cosmological evolution (see section~\ref{sec:cosmo_evolution}). 
In fact, $\phi$ decays while being semi- or ultra-relativistic up to lifetimes of $\tau_\phi \lesssim 10^5\,$s.
Therefore, the total energy density and correspondingly the upper bound on $(n_\phi/n_\gamma)_{T = T^{\text{cd}}}$ depends on the temperature of $\phi$ at chemical decoupling, $T_\phi^\text{cd}$. For a decoupled dark sector this temperature does not need to be the same as the photon temperature $T^\text{cd}$, and hence we show different bounds for different assumed temperature ratios. As expected, the bounds are stronger for a dark sector which is hotter than the SM sector, as then the overall energy density is larger. For $\tau_\phi \gtrsim 10^5\,$s, $\phi$ becomes non-relativistic before its decay and the energy density only depends on the number density, which is why all the different 
curves merge at around this lifetime. Again, as in the case of $m_\phi=60 \,$MeV, we observe the limit to strengthen with the lifetime, although for a given lifetime $\tau_\phi$ the bound on the abundance is considerably weaker, as $\phi$ gets non-relativistic only much later, resulting in a smaller abundance prior to its decay.

\begin{figure}
\begin{center}
\hspace*{-0.7cm}
\includegraphics[scale=0.68]{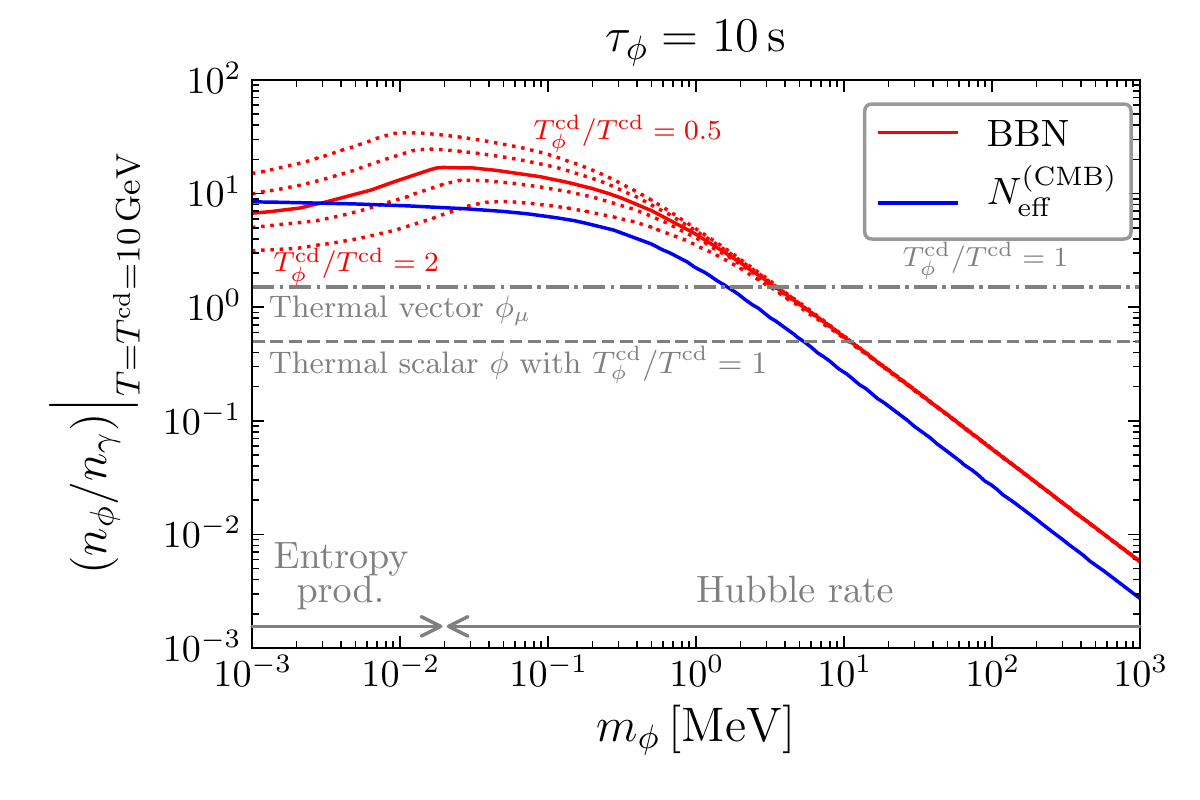}\hspace*{-0.2cm}
\includegraphics[scale=0.68]{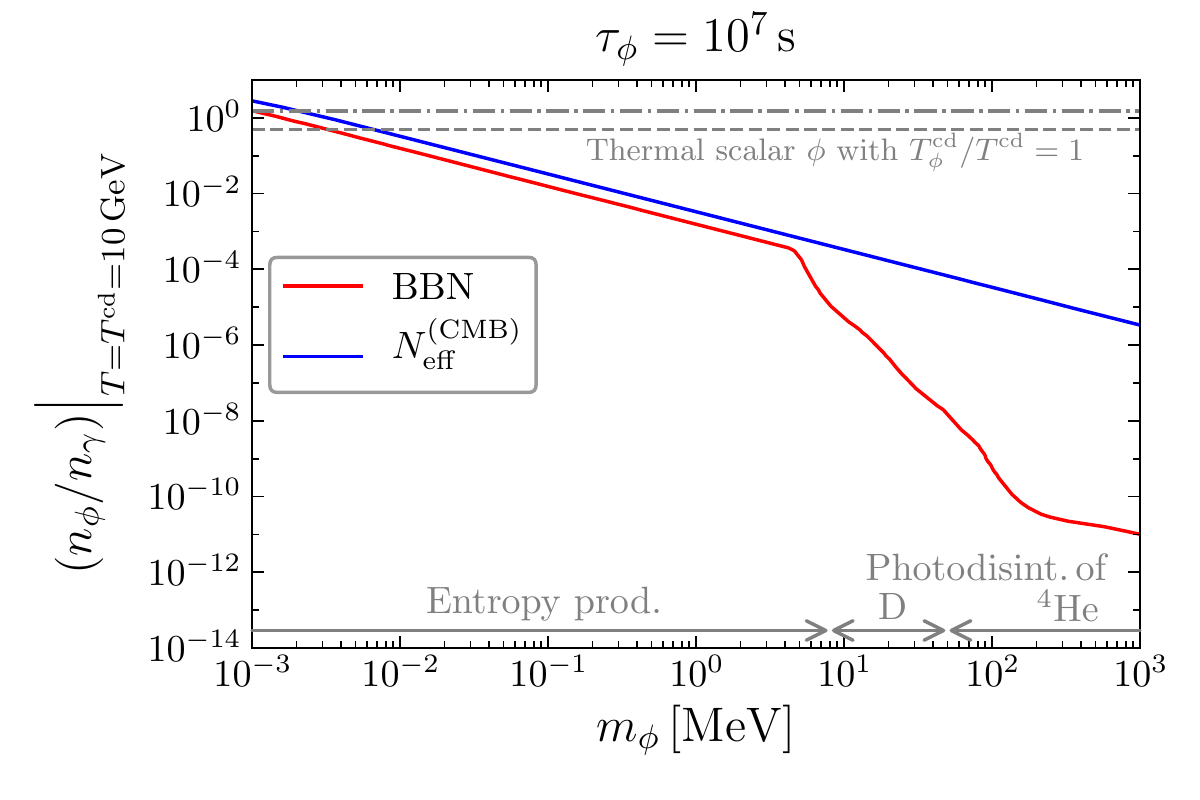}
\end{center}
\caption{Constraints on the abundance of $\phi$ as a function of $m_\phi$ for $\tau_\phi = 10\,$s (left) and $\tau_\phi = 10^7\,$s (right). As in figure~\ref{fig:results_fixed_mphi}, the blue and red curves correspond to the $2\sigma$ upper limits from BBN and $N_\text{eff}^\text{(CMB)}$, respectively, while the grey lines indicate the abundances expected for a thermal vector or scalar particle.}
\label{fig:results_fixed_tauphi}
\end{figure}
In figure~\ref{fig:results_fixed_tauphi} we show the upper bounds on the abundance of $\phi$  as a function of $m_\phi$ for fixed lifetimes $\tau_\phi = 10\,$s (left) and $\tau_\phi = 10^7\,$s (right). In the former case, photodisintegration is irrelevant. Instead, for $m_\phi \gtrsim 0.02\,$MeV, the bound dominantly arises from the increased Hubble rate during BBN.
Up to $m_\phi \simeq 0.1\,$MeV, the bound remains approximately flat as for smaller masses $\phi$ decays while being relativistic. 
For larger values of $m_\phi$ the bound becomes stronger due to the non-relativistic scaling of the particle prior to its decay (see discussion above). 
For $m_\phi \lesssim 0.02\,$MeV on the other hand, the Lorentz boost in the decay of $\phi$ is so large that the actual lifetime is increased from its proper value $\tau_\phi = 10\,$s to $\gtrsim 200\,$s. Hence, the bounds start to get stronger for very small values of $m_\phi$, as more and more entropy is injected between the formation of light elements at $\simeq 180\,$s and recombination. Again, for relativistic decays the kinetic energy and hence the temperature at chemical decoupling $T_\phi^\text{cd}$ is relevant.

For a much larger lifetime, $\tau_\phi = 10^7\,$s, $\phi$ is non-relativistic during its decay for all values of $m_\phi$ considered in the plot.
For masses $m_\phi \lesssim 10\,$MeV, the decay of $\phi$ and the subsequent electromagnetic cascade process only leads to photons with an energy below $E_\gamma \simeq 4\,$MeV, implying that the photons cannot efficiently disintegrate any light nuclei. Hence, for this range of masses the dominant constraints arise from entropy production between BBN and recombination, c.f.~the discussion above.
Once $m_\phi$ is large enough, the bound from photodisintegration is much stronger than the one from entropy production.

In all of the previously discussed plots, the (photon) temperature at which $\phi$ chemically decouples was fixed to the benchmark value $T^\text{cd} = 10\,$GeV. Below this temperature, $\phi$ is only subject to redshift (and decay), while the SM bath undergoes the QCD phase transition at $T\sim 100\,$MeV and thus cools more slowly. This decreases the relative contribution of the energy density of $\phi$. If the decoupling temperature was at a value below the QCD phase transition, $T^\text{cd-alt.} \lesssim 100\,$MeV, the upper bound on the abundance of $\phi$ will be stronger. It is straightforward to see that for particles which decay while being non-relativistic, the bound on $(n_\phi/n_\gamma)_{T = T^\text{cd}}$ then simply scales with a factor $g_s(T^\text{cd-alt.})/g_s(T^\text{cd})$. In addition, our results can also be generalised to other values of the decoupling temperature in the regime where $\phi$ decays while being semi- or ultra-relativistic: In this case, the bound corresponding to a given temperature ratio $T_\phi^\text{cd-alt.}/T^\text{cd-alt.}$ follows from taking the bound for our benchmark choice  $T^\text{cd} = 10\,$GeV with a temperature ratio
\begin{align}
\left( \frac{T_\phi}{T} \right)_{T^\text{cd}} = \left( \frac{g_s(T^\text{cd})}{g_s(T^\text{cd-alt.})} \right)^{1/3} \times \left( \frac{T_\phi}{T} \right)_{T^\text{cd-alt.}}\;\,,
\end{align}
together with an additional scaling $g_s(T^\text{cd-alt.})/g_s(T^\text{cd})$ as explained above.

\subsection{Constraints for particles with thermal abundance}

\begin{figure}
\begin{center}
\hspace*{-0.7cm}
\includegraphics[scale=0.68]{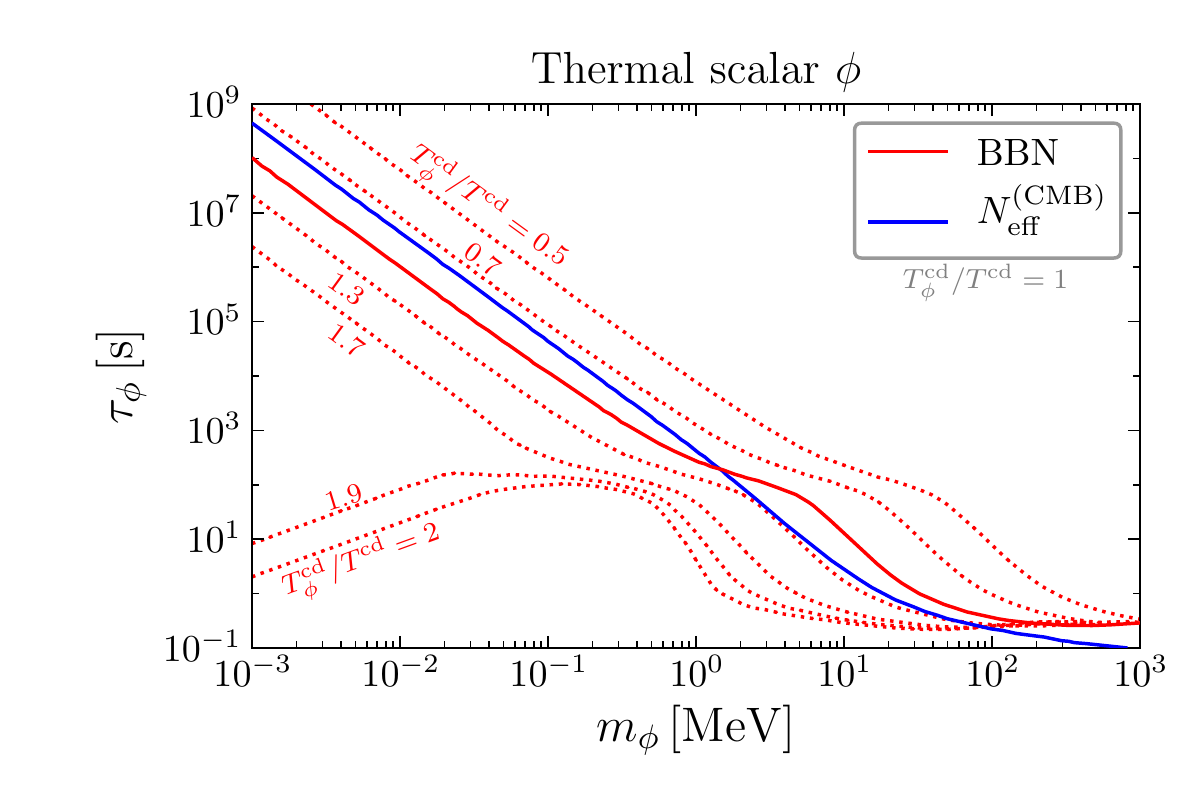}\hspace*{-0.2cm}
\includegraphics[scale=0.68]{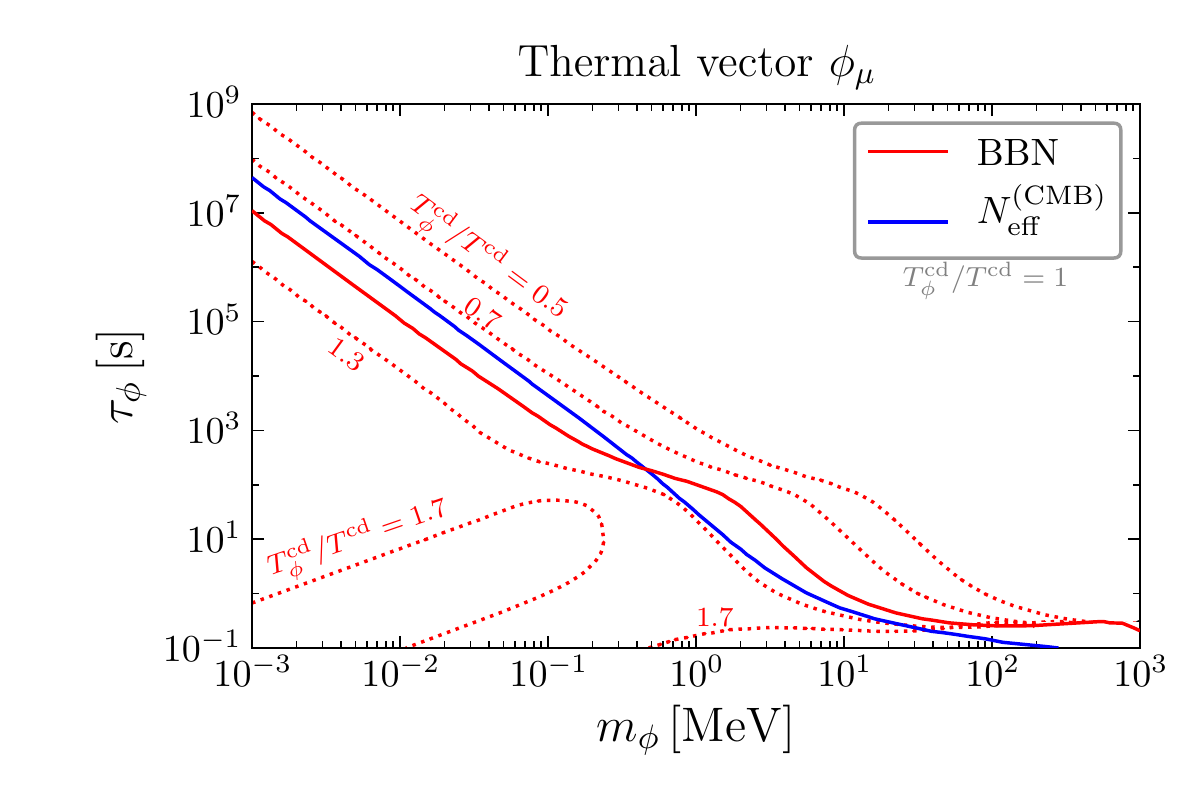}
\end{center}
\caption{Regions of parameter space excluded at $2\sigma$ by BBN (red) and $N_\text{eff}^\text{(CMB)}$ (blue), fixing the abundance of the particle $\phi$ to the one expected for a scalar (left panel) or vector (right panel) with thermal abundance at $T^\text{cd} = 10\,$GeV. The BBN bounds are shown for different temperature ratios $T_\phi^\text{cd}/T^\text{cd}$, while the constraint from $N_\text{eff}^\text{(CMB)}$ is only shown for $T_\phi^\text{cd}/T^\text{cd} = 1$.}
\label{fig:summaryplot_nphithermal}
\end{figure}

Finally, in figure~\ref{fig:summaryplot_nphithermal} we present our results for arbitrary combinations of the mass $m_\phi$ and the lifetime $\tau_\phi$, thereby fixing the initial abundance of the particle $\phi$ to the one expected for a thermally produced scalar (left) or vector (right).
While the limits from photodisintegration are considerably stronger, the bounds shown in figure~\ref{fig:summaryplot_nphithermal} are entirely determined by the constraint on either the increased Hubble rate or the entropy production as photodisintegration is only important for considerably smaller abundances (see Figs.~\ref{fig:results_fixed_mphi} and~\ref{fig:results_fixed_tauphi}).
As expected, the bounds are stronger for a thermally produced vector particle compared to the case of a scalar and are also stronger for larger temperature ratios $T_\phi^\text{cd}/T^\text{cd}$. The bound on $N_\text{eff}^\text{(CMB)}$, shown only for the benchmark case $T_\phi^\text{cd}/T^\text{cd} = 1$, gives a constraint comparable to the one from BBN.

For lifetimes $\tau_\phi \lesssim 200\,$s, the shape of the exclusion bound depends sensitively on whether the particle $\phi$ decays non-relativistically or not. For $m_\phi \gtrsim 1\,$MeV and for the lifetimes relevant for BBN, the decay happens while $\phi$ is semi- or non-relativistic. In this case, the dominant bound for $\tau_\phi \lesssim 200\,$s arises from the increased Hubble rate during BBN (see discussion above). As this constraint is slightly weaker than the one from entropy production, which is relevant for larger lifetimes, the constraints get slightly less stringent for small $\tau_\phi$, which explains the change of slope in the exclusion boundaries at $\tau_\phi \simeq 200\,$s. On the other hand, for $m_\phi \lesssim 1\,$MeV and sufficiently small lifetimes, $\phi$ decays while being ultra-relativistic. As shown already in the left panel of figure~\ref{fig:results_fixed_tauphi}, in this case the Lorentz boost substantially delays the decay of $\phi$, meaning that the entropy production becomes again relevant for very small masses. This is why e.g.~in the left panel of figure~\ref{fig:summaryplot_nphithermal} the exclusion boundary for a temperature ratio $T_\phi^\text{cd}/T^\text{cd} = 2$ starts to bend over at $m_\phi \simeq 10^{-1}\,$MeV. In fact, in this region of parameter space, the BBN bound is stronger for \emph{smaller} values of $m_\phi$ as this enlarges the actual decay time of $\phi$ and thus leads to a larger production of entropy.

For $\tau_\phi \gtrsim 200\,$s, the additional production of entropy after BBN is always relevant. For larger $m_\phi$ and/or $\tau_\phi$, the particle $\phi$ profits longer from the non-relativistic scaling of its energy density, which explains why, in this region of parameter space, the minimal value of the lifetime excluded by BBN becomes larger for smaller masses $m_\phi$. For example, for a scalar particle with mass $m_\phi = 10^{-2}\,$MeV and a temperature ratio $T_\phi^\text{cd}/T^\text{cd} = 1$ at decoupling, only lifetimes $\tau_\phi \gtrsim 10^6\,$s are excluded by BBN, in clear contrast to the often adopted `na\"ive' limit of $1\,$s.

\section{Application to a model of self-interacting dark matter}
\label{sec:SIDM}
After this general discussion of BBN bounds on MeV-scale particles $\phi$ that decay into electron-positron pairs or photons, let us now explore the implications in case this 
particle additionally couples to dark matter, which we denote by $\psi$ in the following. In this setup, the couplings between the SM and the dark sector states are typically constrained to be very small, implying a small annihilation cross section and typically a DM relic abundance which is larger than the observed value of $\Omega_\psi h^2 \simeq 0.12$~\cite{Aghanim:2018eyx}.
The DM relic density can however naturally be achieved if $\psi$ is heavier than $\phi$, $m_\psi \gg m_\phi$, as in this case DM annihilations into a pair of mediators
are kinematically possible, $\psi\bar\psi \rightarrow \phi \phi$, and thermal freeze out can proceed within the dark sector itself. 
However, to ensure a cosmologically viable model, the mediators have to disappear before they dominate the energy density of the universe.
This might be achieved via small couplings to lighter states (which in the simplest setups belong to the SM), rendering these mediators unstable. A different option would be to allow for efficient annihilation of the mediators into even lighter
dark sector states~\cite{Ma:2017ucp,Duerr:2018mbd}.

Interestingly, for $m_\psi \gg m_\phi$, large DM self-interactions are also
naturally present, which may be desirable as they have been argued to alleviate possible small scale tensions of the $\Lambda$CDM paradigm. In particular, the self-scatterings due to $\phi$ exchange are velocity-dependent over large parts of the parameter space
and increase towards smaller velocity, rendering the strong limits on the self-scattering cross section from large velocity systems such as galaxy clusters harmless~\cite{Markevitch:2003at,Randall:2007ph,Peter:2012jh,Rocha:2012jg,Kahlhoefer:2013dca,Harvey:2015hha,Kaplinghat:2015aga}. Nevertheless there are strong bounds on such a setup from {\it (i) dark matter direct detection experiments}, {\it (ii) the CMB} and {\it (iii)} presumably {\it BBN} which is the
focus of this study. The interplay of these different constraints depends in particular on the quantum numbers and coupling structure of the mediators. For the case of $s$-wave DM annihilation into the mediator and subsequent mediator decays into SM states such as electrons and photons, there are very strong reionisation bounds from the CMB and the parameter space leading to interesting dark matter self-scattering cross sections is essentially excluded~\cite{Bringmann:2016din,Cirelli:2016rnw}. This observation applies in particular for vector mediators that are kinetically mixed with the SM. Scalar mediators on the other hand lead to $p$-wave annihilation and the overwhelming CMB bounds do not apply. Scalars naturally couple via the Higgs portal, implying a Yukawa-like coupling structure and hence rather small couplings to electrons. For mediator masses below $m_\phi \lesssim 200$~MeV, which are relevant for self-interacting DM, the mediator lifetime can therefore be sizeable and  bounds from BBN are expected to be very relevant in this context~\cite{Kaplinghat:2013yxa,Kainulainen:2015sva,Kahlhoefer:2017umn}.

\subsection{A simple model of DM self-interactions}
\label{sec:SIDM_model}

In the following we study a very simple model featuring a dark matter particle $\psi$ and a scalar mediator $\phi$, which couples to the SM via a Higgs portal coupling, with the relevant terms in the Lagrangian given by
\begin{equation}
\mathcal{L}_\phi \supset - \,  y_\psi  \, \bar \psi \psi \phi - \, y_{\text{SM}} \sum_f  \frac{m_f}{\vev} \, \, \bar f f \phi \;\,.
\label{eq:LDM}
\end{equation}
Here, $\vev \simeq 246\,$GeV is the electroweak vacuum expectation value and $m_f$ is the mass of the SM fermion $f$. This model has been extensively studied and is known to exhibit large DM self-interactions in large regions of the parameter space~\cite{Kaplinghat:2013yxa,Kainulainen:2015sva,Kahlhoefer:2017umn}. A detailed discussion of the corresponding momentum transfer cross section $\sigma_\text{T}$ within this model can be found in~\cite{Kahlhoefer:2017umn}, including effects arising from the indistinguishability of the scattered particles, hence we will not repeat this discussion here. 

The relic abundance of the DM particle is determined by hidden sector freeze out via the process $\psi\bar\psi \rightarrow \phi \phi$. As we will discuss below, experimental constraints on the SM coupling $y_{\text{SM}}$ are often so stringent that the two sectors \textcolor{black}{cease} to be in thermal equilibrium at temperatures much larger than the freeze out temperature, which can substantially affect the standard calculation of the dark matter relic density. To take this into account, we determine the relic density $\Omega_\psi h^2$ for given values of the masses $m_\psi, \,m_\phi$ and the couplings $y_\psi, y_{\text{SM}}$ as follows: First we determine the smallest temperature $T^\text{vd}$ at which the two sectors are still in thermal equilibrium, which is given by $\Gamma_{\psi \bar \psi \to f \bar f}(T^\text{vd}) = H(T^\text{vd})$.\footnote{For $T \lesssim 5\,$GeV, where the light SM quarks are no longer the appropriate degrees of freedom, the cross section for $\psi \bar \psi \to f \bar f$ can be expressed in terms of the width of a hypothetical scalar particle with mass $m = \sqrt{s}$~\cite{Cline:2013gha,Alekhin:2015byh}.} For $T < T^\text{vd}$, the temperature of the dark sector $T_\phi$ in general deviates from the photon temperature $T$, and we determine $T_\phi(T) \big|_{T < T^\text{vd}}$ by demanding separate entropy conservation in both thermal baths. The actual freeze-out of the dark matter particle via the annihilation into mediators then occurs at a (photon) temperature $T^\text{cd}$ defined via $\Gamma_{\psi \bar \psi \to \phi \phi}(T_\phi^\text{cd}) = H(T^\text{cd})$. Following~\cite{Cassel:2009wt,Iengo:2009ni,Slatyer:2009vg}, in the calculation of the corresponding annihilation rate $\Gamma_{\psi \bar \psi \to \phi \phi}$ we take into account Sommerfeld enhancement associated to the multiple exchange of mediators in the initial state, which can be important for a sufficiently large hierarchy $m_\psi \gg m_\phi$, even for the rather large velocities during thermal freeze out. We then finally obtain the relic density $\Omega_\psi h^2$ by assuming that the yield $Y_\psi = n_\psi/s$ stayed constant between $T^\text{cd}$ and today. For the case of standard s-wave freeze-out in the visible sector, we checked that this approximate way of computing the dark matter relic density via the assumption of instantaneous freeze-out is in $\mathcal{O}(10\,\%)$ agreement with the full numerical solution of the Boltzmann equation~\cite{Steigman:2012nb}, which is accurate enough for our purposes.

In our analysis, we then determine the value of the DM-mediator coupling $y_\psi$ giving rise to the observed abundance of dark matter $\Omega_\psi h^2 = 0.12$, which is in general a function of $m_\psi$, $m_\phi$ and $y_\text{SM}$. However, for sufficiently small values of $y_\text{SM}$, the dark and the visible sector have actually \emph{never} thermalised via the annihilation process $\psi \bar \psi \to f \bar f$, i.e.~there is no value $T^\text{vd}$ satisfying $\Gamma_{\psi \bar \psi \to f \bar f}(T^\text{vd}) = H(T^\text{vd})$. For $m_\psi \lesssim m_t$, we find that this occurs whenever $y_\text{SM} \, y_\psi \lesssim 1.1 \times 10^{-6}$ (see also~\cite{Kahlhoefer:2017umn}). In this case, there is a priori no direct link between the temperatures of both sectors, which renders the model considerably less predictive. However, in this case it is very conceivable that both sectors have been equilibrated by some other high-scale interactions which subsequently froze out, giving rise to a temperature ratio of the two sectors of $\mathcal{O}(1)$. In those parts of the parameter space, for simplicity we will thus assume that $T_\phi^\text{vd} \equiv T^\text{vd}$, with $T^\text{vd}$ being determined for the smallest value of $y_\text{SM} \, y_\psi$ for which both sectors have still equilibrated, which typically happens at $T^\text{vd} \simeq \text{max}(m_\psi, m_t)$~\cite{Kahlhoefer:2017umn}. Nevertheless, we will indicate in which regions of parameter space this assumption is necessary and where the corresponding constraints from BBN might be considerably weakened for smaller temperature ratios.

\subsection{Dark matter direct detection and other constraints}
\label{sec:SIDM_DDAndOther}

Given that the mediator mass of interest is sub-GeV, it is unsurprising that the coupling $y_{\text{SM}}$ to SM states is strongly constrained (see e.g.~\cite{Krnjaic:2015mbs,Alekhin:2015byh}).
For the mediator masses of interest, the strongest upper bounds on $y_{\text{SM}}$ typically come from searches for rare kaon decays as described in~\cite{Kahlhoefer:2017umn}, leading to
\begin{align}
y_\text{SM} \lesssim 1.9 \cdot 10^{-4}  \,\; .
\end{align} 

In addition, there are strong constraints from astrophysical observations, for example stemming from an analysis of the SN1987a supernova neutrinos~\cite{Krnjaic:2015mbs} or the lifetime of horizontal branch stars~\cite{Raffelt:1987yu}. The latter are only relevant for mediator masses $m_\phi < 30\,\mathrm{keV}$, but the supernova bound may be relevant for large regions of the parameter space although it still suffers from significant theoretical uncertainties.

Dark matter direct detection experiments constrain the coupling combination $y_\psi \cdot y_{\text{SM}}$ as can be seen from the spin-independent DM scattering rate on nuclei~\cite{Kahlhoefer:2017umn},
\begin{equation}
 \frac{\text{d}\sigma^\text{SI}_T}{\text{d}E_R} = \frac{f_p^2 \, m_p^2}{2 \pi \, \vev^2} \frac{m_T A^2 F^2(E_R)}{v^2} \frac{y_\psi^2 \, y_\text{SM}^2}{(m_\phi^2 + q^2)^2} \;\, .
\label{eq:sigN}
\end{equation}
Here $f_p = f_n \approx 0.3$ is the effective nucleon coupling~\cite{Cline:2013gha}, $F^2(E_R)$ the form factor for spin-independent scattering, $v$ the DM velocity and $q$ the transferred momentum in a nuclear recoil event.
As the coupling $y_\psi$ is basically fixed by the requirement of achieving the observed DM relic abundance, this effectively translates into a bound on $y_{\text{SM}}$. The strongest constraints on the scattering rate are set by XENON1T~\cite{Aprile:2018dbl} for large and by CRESST-II~\cite{Angloher:2015ewa,Angloher:2017zkf} and CDMSlite~\cite{Agnese:2015nto} for smaller DM masses. We evaluate these bounds by employing the publically available code \textsc{DDCalc 2.0.0}~\cite{Workgroup:2017lvb,Athron:2018hpc}, which we modified in order to take into account the non-standard dependence of the recoil rate on the momentum transfer $q^2$ due to the presence of the light mediator $\phi$. We also study to what extent future direct detection experiments might further probe the parameter space of the model; specifically, we consider the final stage of the CRESST-III experiment~\cite{Angloher:2015eza} which plans to achieve an exposure of $\simeq 1000\,\text{kg}\,\text{days}$ and a threshold of $\simeq 100\,$eV. Details of our implementation of CRESST-III can be found in~\cite{Kahlhoefer:2017ddj}.

Lastly, for very small couplings $y_\text{SM}$, the lifetime of the mediator $\phi$ can become so large that the electromagnetic decay products do not thermalise with the background photons, thus giving rise to spectral distortions in the CMB. This excludes all relevant parameter space with $\tau_\phi \gtrsim 10^8\,$s~\cite{Poulin:2016anj}; as we will see in the next section, a more precise calculation of the bound on $\tau_\phi$ (which in general depends on the abundance of $\phi$ and thus on the masses and couplings of the model) will not affect any of our conclusions regarding the viability of self-interactions via a scalar mediator, meaning that this simple estimate is sufficient for our purposes.


\subsection{Resulting constraints from BBN}
\begin{figure}
\begin{center}
\hspace*{-0.6cm}
\includegraphics[scale=1.2]{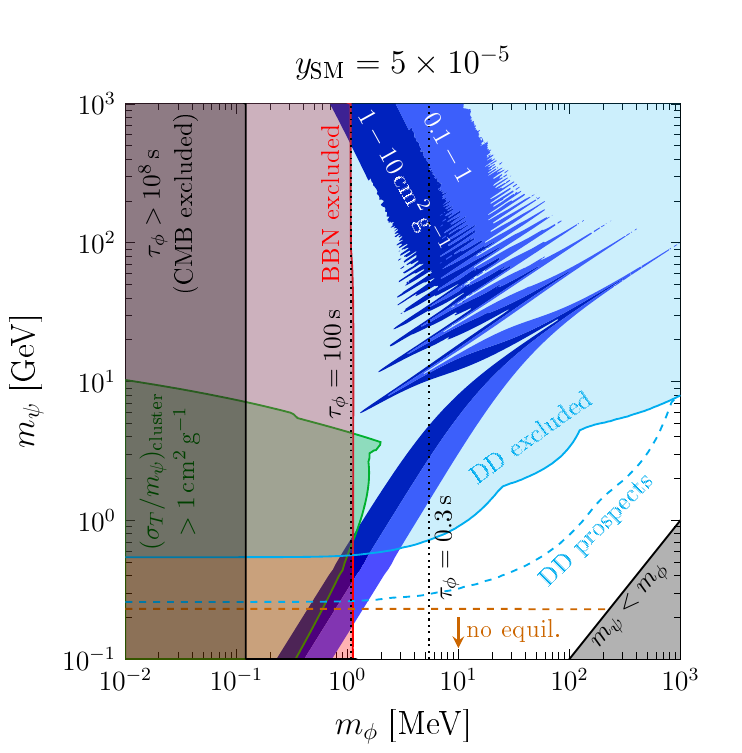}
\end{center}
\caption{Constraints on the parameter space of self-interacting dark matter with mass $m_\psi$ coupled to a mediator with mass $m_\phi$. The coupling $y_\psi$ between these two particles is chosen at each point in parameter space from the requirement that $\Omega_\psi h^2 \simeq 0.12$, while the coupling of $\phi$ to SM states is fixed to $y_\text{SM} = 5 \times 10^{-5}$. The dark and light blue shaded regions correspond to a self-interaction cross section on dwarf scales of $1\,\text{cm}^2/\text{g} < \sigma_T/m_\psi < 10\,\text{cm}^2/\text{g}$ and  $0.1\,\text{cm}^2/\text{g} < \sigma_T/m_\psi < 1\,\text{cm}^2/\text{g}$, respectively, while the green shaded region shows the bound from cluster observations. Direct detection excludes everything inside the cyan shaded part of parameter space, with the future CRESST-III experiment being potentially able to push this limit down to the dashed cyan curve. The bounds from BBN and CMB are shown in red and dark grey, respectively.}
\label{fig:SIDM_Fixed_ySM5e-5}
\end{figure}

In figure~\ref{fig:SIDM_Fixed_ySM5e-5} we present the various constraints on the model discussed in the previous section for a fixed choice of the coupling $y_\text{SM} = 5 \times 10^{-5}$, with the mediator mass $m_\phi$ on the horizontal and the DM mass $m_\psi$ on the vertical axis. As explained above, the DM-mediator coupling $y_\psi$ is fixed for each point in parameter space by the requirement of correctly reproducing the observed relic density of DM. In the dark and light blue shaded regions, the momentum transfer cross section of DM at a velocity of $30 \,$km/s (a typical velocity at small scales) lies within $1\,\text{cm}^2/\text{g} < \sigma_T/m_\psi < 10\,\text{cm}^2/\text{g}$ and  $0.1\,\text{cm}^2/\text{g} < \sigma_T/m_\psi < 1\,\text{cm}^2/\text{g}$, respectively, which is roughly the range required for addressing the small-scale problems of the $\Lambda$CDM model (see e.g.~\cite{Vogelsberger:2012ku,Tulin:2013teo,Tulin:2017ara}). For dark matter masses $m_\psi \gtrsim 0.6\,$MeV, a large part of the parameter space is excluded by direct detection experiments, shown by the cyan shaded region. On the other hand, the red shaded region indicates which combinations of $m_\phi$ and $m_\psi$ are excluded by BBN. The corresponding calculation follows directly from the discussion in sections~\ref{sec:cosmo_evolution} and~\ref{sec:abundance_of_light_elements}; the lifetime $\tau_\phi$ and branching ratios into the different final states are taken from~\cite{Bezrukov:2009yw,Fradette:2017sdd}. Evidently, for this particular choice of $y_\text{SM}$ the combination of direct detection and BBN constraints excludes almost all parameter space with sufficiently large self-interaction cross section of DM on the scale of dwarf galaxies. Additional constraints from the CMB (grey shaded) as discussed in the previous section, as well as the upper bound $\sigma_T/m_\psi < 1\,\text{cm}^2/\text{g}$ on the scale of galaxy clusters ($v \simeq 1000\,$km/s) shown in green do not impose further restrictions on the viable range of parameters. However, it is interesting to note that CRESST-III will be able to probe all of the remaining parameter space leading to $1\,\text{cm}^2/\text{g} < \sigma_T/m_\psi < 10\,\text{cm}^2/\text{g}$, as indicated by the dashed cyan curve. Finally, let us remark that as long as $m_\psi \gtrsim 120\,$MeV (shown by the orange dashed curve), the chosen value for $y_\text{SM}$ is large enough such that the dark and visible sector have been in equilibrium at high temperatures.

\begin{figure}
\begin{center}
\hspace*{-0.7cm}
\includegraphics[scale=0.9]{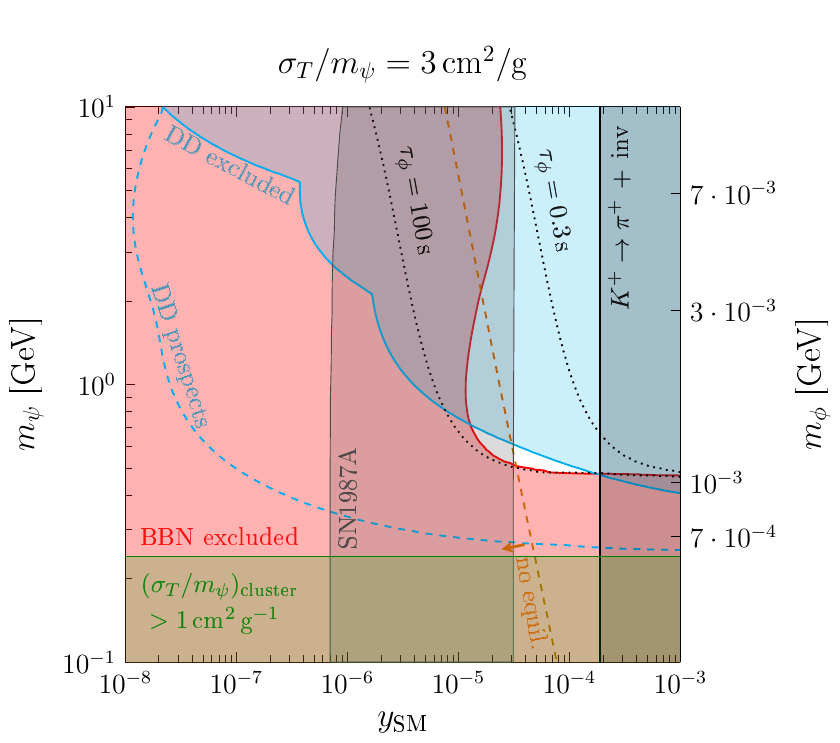}\hspace{0.2cm}
\includegraphics[scale=0.9]{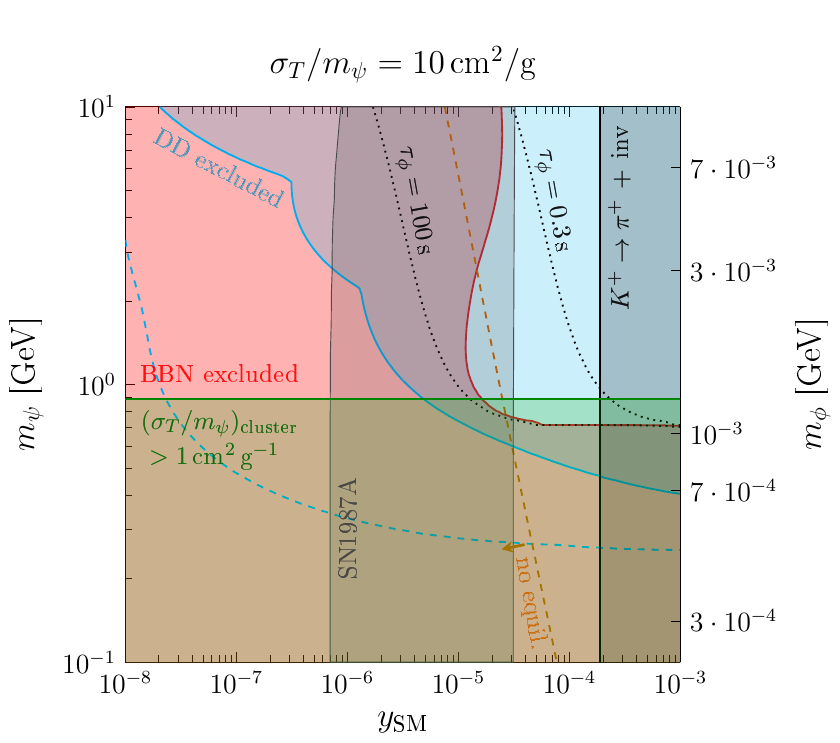}
\end{center}
\caption{Constraints for a fixed self-interaction cross section $\sigma_T/m_\psi = 3\,\text{cm}^2/\text{g}$ (left panel) and $\sigma_T/m_\psi = 10\,\text{cm}^2/\text{g}$ (right panel). In addition to the bounds already shown in figure~\ref{fig:SIDM_Fixed_ySM5e-5}, we also show the constraints from SN1987A (adapted from \cite{Krnjaic:2015mbs}) and rare kaon decays in grey. Values of $y_\text{SM}$ below the dashed orange curve correspond to scenarios in which the interaction of the mediator $\phi$ alone is not sufficiently strong in order to equilibrate the dark and visible sectors at high temperatures.}
\label{fig:SIDM_Fixed_sigmaT}
\end{figure}

While the region of parameter space leading to the desired self-interaction cross section of DM is (nearly) insensitive to the choice of $y_\text{SM}$, both the constraints from direct detection and from BBN depend strongly on the particular choice employed in figure~\ref{fig:SIDM_Fixed_ySM5e-5}. In order to investigate for which values of $y_\text{SM}$ one has viable regions with strong self-interactions of DM, we show in figure~\ref{fig:SIDM_Fixed_sigmaT} the relevant bounds as a function of the coupling $y_\text{SM}$ and the DM mass $m_\psi$, fixing the mediator mass $m_\phi$ such that at each point in parameter space the momentum transfer cross section of DM for $v=30\,$km/s  is equal to $3\,\text{cm}^2/\text{g}$ (left panel) or $10\,\text{cm}^2/\text{g}$ (right panel). As can be seen from figure~\ref{fig:SIDM_Fixed_ySM5e-5}, this construction is possible in the region of interest, $m_\psi \lesssim 10\,$GeV, in which case there is a unique choice of $m_\phi$ leading to a given value of $\sigma_T/m_\psi$. In addition to the bounds already depicted in the previous figure, here we also show which values of $y_\text{SM}$ are excluded by rare kaon decays (dark grey shaded region) or are disfavoured by the SN1987A observation (light grey shaded region), following the discussions in section~\ref{sec:SIDM_DDAndOther}. Figure~\ref{fig:SIDM_Fixed_sigmaT} gives a clear view on the strong complementarity of the bounds from direct detection experiments and rare kaon decays on the one hand, and from BBN on the other hand. The former are relevant for sufficiently large values of the coupling $y_\text{SM}$, with direct detection being more and more sensitive for larger DM masses, while BBN excludes regions of parameter space with small values of $y_\text{SM}$, corresponding to large lifetimes $\tau_\phi$. The region $m_\phi < 2 m_e$ is ruled out by BBN for all values of $y_\text{SM}$ shown in figure~\ref{fig:SIDM_Fixed_sigmaT}, due to the strongly suppressed decay width of $\phi \rightarrow \gamma \gamma$, resulting in large values of the lifetime $\tau_\phi$.

Taken together, all the constraints only leave a narrow window of viable parameter space leading to $\sigma_T/m_\psi = 3\,\text{cm}^2/\text{g}$, centred around $m_\psi \simeq 0.5\,$GeV, $m_\phi \simeq 1.1\,$MeV and $y_\text{SM} \simeq 5 \times 10^{-5}$ (see also figure~\ref{fig:SIDM_Fixed_ySM5e-5}). Note that this combination of model parameters corresponds to a lifetime $\tau_\phi \simeq 30\,\text{s} \gg 1\,\text{s}$, but is nevertheless \emph{not} excluded by BBN. Remarkably, future low-threshold direct detection experiments such as CRESST-III will be able to fully probe the remaining parameter space, as indicated by the dashed cyan curve. On the other hand, the right panel of figure~\ref{fig:SIDM_Fixed_sigmaT} shows that an even larger self-interaction cross section of $\sigma_T/m_\psi = 10\,\text{cm}^2/\text{g}$ is now already robustly excluded by the combination of direct detection experiments and BBN, for all values of the coupling $y_\text{SM}$.\footnote{In particular, this conclusion does not rely on the bound from SN1987A, which suffers from significant systematic uncertainties.} However, it is important to note that for sufficiently small values of $y_\text{SM}$ the dark and visible sector have never been in thermal contact via the exchange of the mediator $\phi$, as shown by the dashed orange curves in both panels of figure~\ref{fig:SIDM_Fixed_sigmaT}. As explained in detail in section~\ref{sec:SIDM_model}, in this part of parameter space we assume for definiteness that $T_\phi^\text{vd}=T^\text{vd}$. 
Allowing for a strongly suppressed temperature of the dark sector would significantly decrease the energy density of $\phi$ and thus lead to less stringent bounds from BBN; a detailed discussion of such a setup is beyond the scope of this work.

\begin{figure}
\begin{center}
\hspace*{-1.2cm}
\includegraphics[scale=0.68, trim = 0 26 0 0]{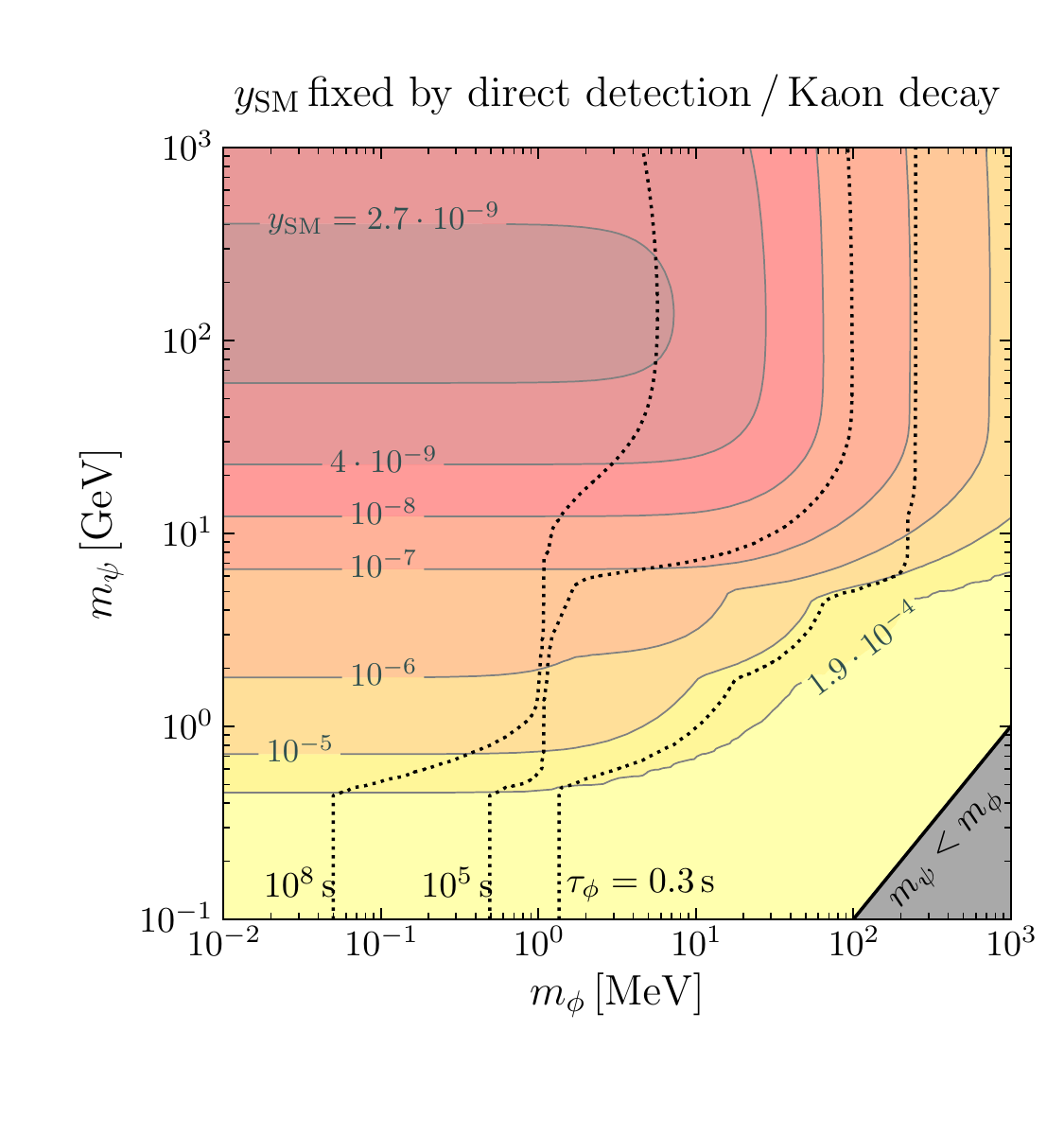}\hspace*{0.0cm}
\includegraphics[scale=1]{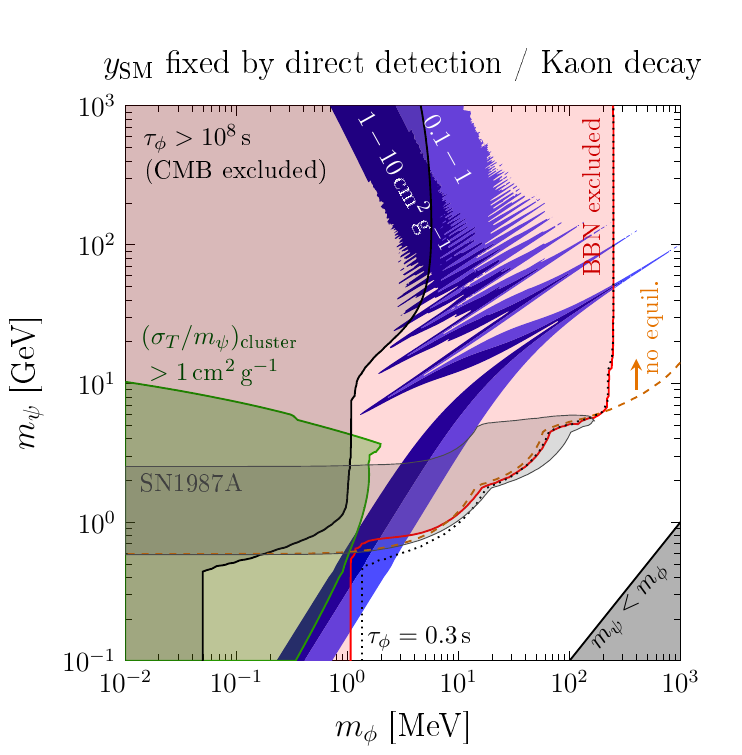}
\end{center}
\caption{Left panel: contours of $y_\text{SM}$ given by the largest possible value allowed from direct detection searches and rare kaon decays. Right panel: corresponding constraints from BBN (red shaded), self-interactions on cluster scales (green shaded), SN1987A and CMB (both gray shaded). Similar as in figure~\ref{fig:SIDM_Fixed_ySM5e-5}, the dark and light blue shaded regions indicate the desired range of the DM self-interaction cross section.}
\label{fig:SIDM}
\end{figure}

Finally, in figure~\ref{fig:SIDM} we present the constraints on the model of self-interacting DM under the assumption that at each point in parameter space the coupling $y_\text{SM}$ takes the largest value which is still compatible with both direct detection searches and rare kaon decays. This choice of $y_\text{SM}$ minimises the lifetime $\tau_\phi$ and thus leads to the least stringent bound from BBN. Hence, in this way one can directly assess the compatibility of direct detection searches, rare kaon decays and nucleosynthesis with the idea of strong self-interactions of DM. First, in the left panel of figure~\ref{fig:SIDM} we show contours of the corresponding values of $y_\text{SM}$. For $m_\psi \lesssim 0.5\,$GeV, direct detection experiments are insensitive and the coupling is fixed to the upper bound $y_\text{SM} = 1.9 \cdot 10^{-4}$ from rare kaon decays (see section~\ref{sec:SIDM_DDAndOther}), while for larger DM masses direct detection provides the stronger constraint. Note that the latter is independent of the mediator mass for $m_\phi \lesssim 1\,$MeV, corresponding to the smallest momentum exchange resolvable by direct detection experiments.

In the right panel of figure~\ref{fig:SIDM}, the red shaded region shows the range of parameters excluded by BBN observations, thereby fixing $y_\text{SM}$ to the values shown in the left panel (and still fixing $y_\psi$ by the requirement that $\Omega_\psi h^2 \simeq 0.12$). Similar as in figure~\ref{fig:SIDM_Fixed_ySM5e-5}, the dark and light blue shaded regions correspond to a self-interaction cross section which is in the interesting range for solving the small-scale problems of standard cold DM, while the green shaded area is disfavoured due to a too large value of $\sigma_T/m_\psi$ on the scale of galaxy clusters. In addition, the dark and light grey shaded regions show exclusion bounds from spectral distortions of the CMB and from observations of SN1987A, as discussed in section~\ref{sec:SIDM_DDAndOther}. Clearly, large parts of the parameter space leading to $0.1\,\text{cm}^2/\text{g} < \sigma_T/m_\psi < 10\,\text{cm}^2/\text{g}$ on dwarf galaxy scales are robustly excluded. Small mediator masses $m_\phi < 2 m_e$ are ruled out by BBN due to the large lifetime of $\phi$ implied by direct detection searches and/or rare kaon decays (see also the dotted curves in the left panel). At larger values of $m_\phi$, the BBN bound becomes less stringent for sufficiently small values of $m_\psi$ due to the insensitivity of direct detection experiments to very small DM masses, allowing for larger values of $y_\text{SM}$ and thus for smaller lifetimes $\tau_\phi$. Finally, for $m_\phi > 2 m_\mu \simeq 210\,$MeV, the mediator can quickly decay into muons, leading to lifetimes well below the onset of BBN ($\tau_\phi \lesssim 0.3\,$s).
Thus, in addition to the remaining part of parameter space with significant self-interactions of DM around $m_\psi \simeq 0.5\,$GeV and $m_\phi \simeq 1.1\,$MeV which has already been discussed in Figs.~\ref{fig:SIDM_Fixed_ySM5e-5} and Figs.~\ref{fig:SIDM_Fixed_sigmaT}, the combinations of $m_\psi$ and $m_\phi$ corresponding to the first and second resonance peak of $\sigma_T/m_\psi$ are also potentially consistent with all existing constraints. A detailed investigation of the viability of this tiny part of parameter space would require a careful consideration of the thermal history of the dark sector taking into account effects associated to the resonant enhancement of the annihilation cross section~\cite{Binder:2017lkj,Bringmann:2018jpr}, which is left for future work.

\section{Conclusions}
\label{sec:conclusions}

Particles with a mass in the MeV range are predicted by various extensions of the Standard Model, with their coupling strength to SM states often being subject to stringent upper limits from direct or indirect searches. Unless one invokes additional even lighter (dark) states into which such a new particle $\phi$ can decay, this generically implies macroscopic lifetimes $\tau_\phi \gtrsim \mathcal{O}(1\,\text{s})$. While it is well known that this can potentially be in conflict with the remarkable success of standard Big Bang Nucleosynthesis, detailed studies of the impact of extra particles on the abundances of light nuclei so far have only been conducted in the limiting cases where the particle is either non-relativistic during BBN, or where it is ultra-relativistic and decays only well after BBN.  However, when considering an MeV-scale particle, the four energy scales set by the particle mass $m_\phi$, the temperature during BBN, the temperature at the time of decay, as well as the binding energy of light nuclei such as deuterium can all be similar, which significantly complicates the physics underlying the calculation of BBN constraints. Motivated by this, we present for the first time a comprehensive study of BBN constraints on MeV-scale particles decaying into $e^+ e^-$ or $\gamma \gamma$ with a lifetime in the range $10^{-2}\,\text{s} < \tau_\phi < 10^8\,\text{s}$.

To this end, we first numerically solve the full Boltzmann equation for the phase-space distribution function of the decaying particle $\phi$, without invoking any ultra- or non-relativistic approximation. Besides the additional energy density of $\phi$, which contributes to the expansion rate of the Universe and thus modifies the primordial abundances, we also consider in detail the production of entropy via the thermalisation of the decay products of $\phi$. This process can lead to a non-standard time-dependence of the baryon-to-photon-ratio $\eta(t)$ during the time when it is most relevant for BBN. We take into account both the enhanced Hubble rate as well as the non-standard baryon-to-photon ratio by properly modifying the public code \textsc{AlterBBN}. We find that in general both effects modify the predicted nuclear abundances in a similar way.
Finally, if $m_\phi \gtrsim 4\,$MeV and $\tau_\phi \gtrsim 10^4\,$s, photodisintegration of light nuclei after the end of BBN can substantially modify the nuclear abundances. As already noted in~\cite{Poulin:2015woa,Poulin:2015opa}, for decaying particles with a mass in the MeV range the usually adopted `universal spectrum' of photons originating from the cascade process on the background photons is typically not applicable. Thus, by fully tracking the cascade evolution of high-energetic photons, electrons and positrons via double photon pair creation, photon-photon scattering, Bethe-Heitler pair creation, Compton and inverse Compton scattering, we derive the non-thermal photon spectrum and the associated photodisintegration rates of deuterium and helium separately for each point in parameter space for a given branching ratio of $\phi$.

We then derive model-independent upper bounds on the ratio of the initial abundance of the particle $\phi$ and the photon-number density $n_\phi/n_\gamma$. To this end, we employ recent data on primordial abundances, and take into account systematic uncertainties on the nuclear rates relevant to BBN. 
Depending on the region in parameter space either of the effects related to the increased Hubble rate, the modified baryon-to-photon ratio or the photodisintegration of light nuclei can dominate the final constraint, reinforcing the necessity of a dedicated study of BBN constraints on MeV-scale particles. Importantly, when fixing the abundance of $\phi$ to the value expected for a thermal relic, we find that our upper limits in large parts of the parameter space deviate significantly from the frequently adopted order-of-magnitude estimates $\tau_\phi \lesssim 1\,$s (corresponding to the start of BBN) or $\tau_\phi \lesssim 10^4\,$s (corresponding to the start of photodisintegration). In appendix~\ref{app:plot_collection} we provide upper limits on $n_\phi/n_\gamma$ for a large set of model parameters, enabling the reader to quickly read off the BBN upper bound on the abundance of an unstable particle decaying into $e^+ e^-$ or $\gamma \gamma$.

Lastly, we apply our general results to a specific model of self-interacting dark matter involving a fermionic dark matter particle $\psi$ interacting with a scalar mediator $\phi$, with the latter also having Higgs-like couplings to SM states. Such a scenario is compelling as it leads to large self-interaction cross sections of dark matter on small scales (and thus potentially solves tensions found within the pure $\Lambda$CDM model), while being consistent with upper bounds on the scales of galaxy clusters. However, the coupling strength of $\phi$ to SM particles is strongly constrained by direct detection experiments, rare kaon decays and bounds from the duration of the neutrino pulse from SN1987A. Based on a careful calculation of the cosmological evolution of both the dark matter particle $\psi$ and the unstable mediator $\phi$, we then derive for the first time detailed BBN constraints on this scenario. Our results show that almost all of the parameter space of the model leading to significant self-interactions of dark matter is ruled out by the combination of direct detection experiments and BBN, with only a small region around $m_\psi \simeq 0.5\,$GeV, $m_\phi \simeq 1.1\,$MeV and $\tau_\phi \simeq 30\,$s remaining. Interestingly, this combination of parameters can be fully tested with upcoming low-threshold direct detection experiments such as the final phase of CRESST-III~\cite{Angloher:2015eza}.\\[0.5cm]
\textit{Note added:}\\
Shortly after the completion of this work, constraints from photodisintegration arising from the decay of MeV-scale particles were also studied in~\cite{Forestell:2018txr}, including for the first time the effect of FSR of photons. While this does not affect our bounds for $\tau_\phi \lesssim 10^4\,$s or $m_\phi \lesssim 5\,$MeV arising from the increased Hubble rate or entropy production (which is not considered in~\cite{Forestell:2018txr}), it does have an impact on the limits for a sufficiently heavy particle decaying into $e^+ e^-$ with a lifetime $\tau_\phi \gg 10^4\,$s. In this updated version of our work, we have thus included FSR via eq.~(\ref{eq:SFSR}), leading to minor changes in the left panel of Fig.~\ref{fig:results_fixed_mphi}, the right panel of Fig.~\ref{fig:results_fixed_tauphi} and the lower left panel of Fig.~\ref{fig:appendix_plots}.

\acknowledgments

We thank Camilo Garcia-Cely and Felix Kahlhoefer for useful discussions. This work is supported by the ERC Starting Grant `NewAve' (638528).

\pagebreak
\appendix

\section{Collection of BBN constraints}
\label{app:plot_collection}
\begin{figure}[h!]
	\begin{center}
		\hspace*{-0.7cm}
		\includegraphics[scale=0.68]{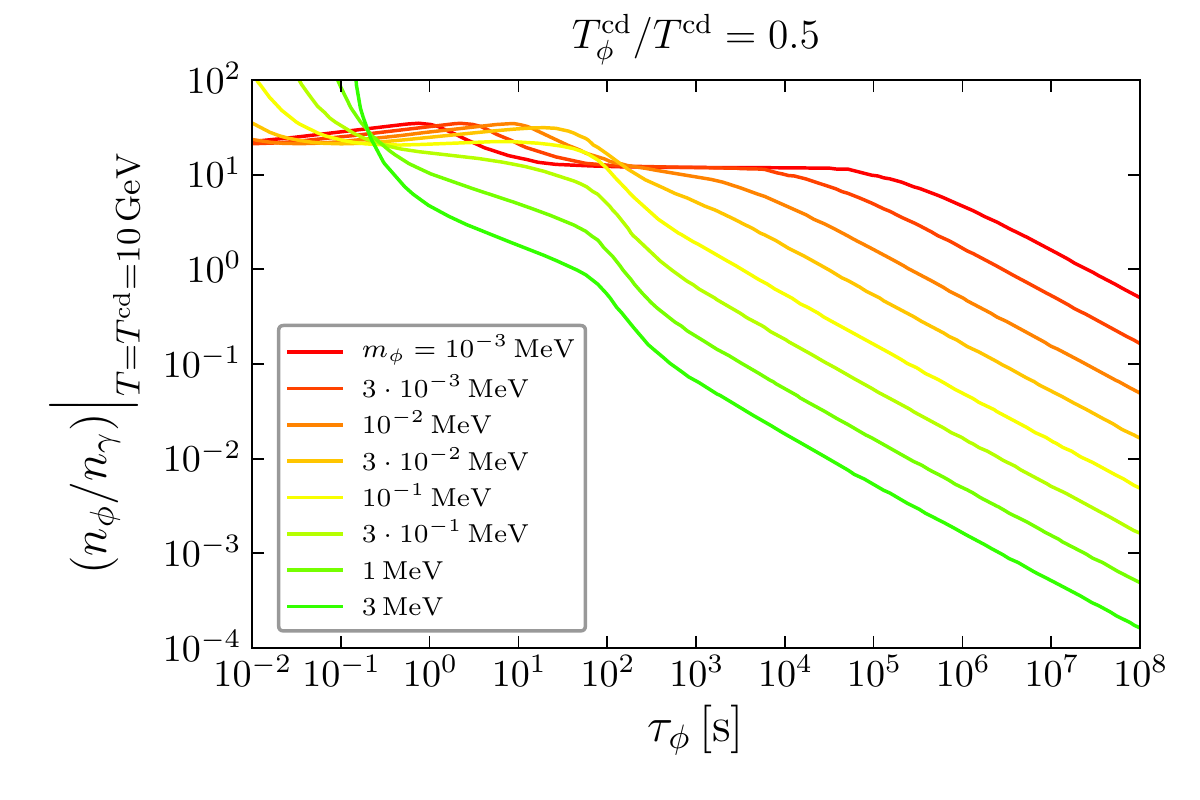}\hspace*{-0.2cm}
		\includegraphics[scale=0.68]{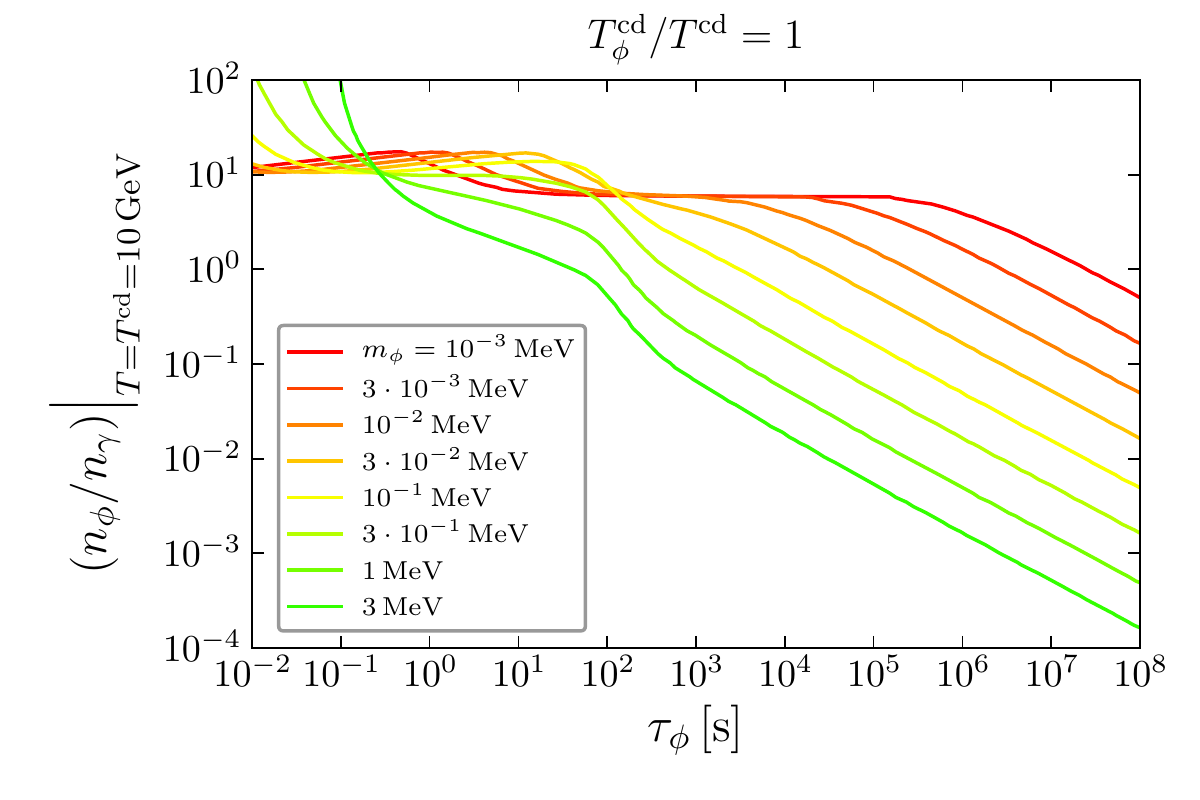}\\
		\hspace*{-0.7cm}
		\includegraphics[scale=0.68]{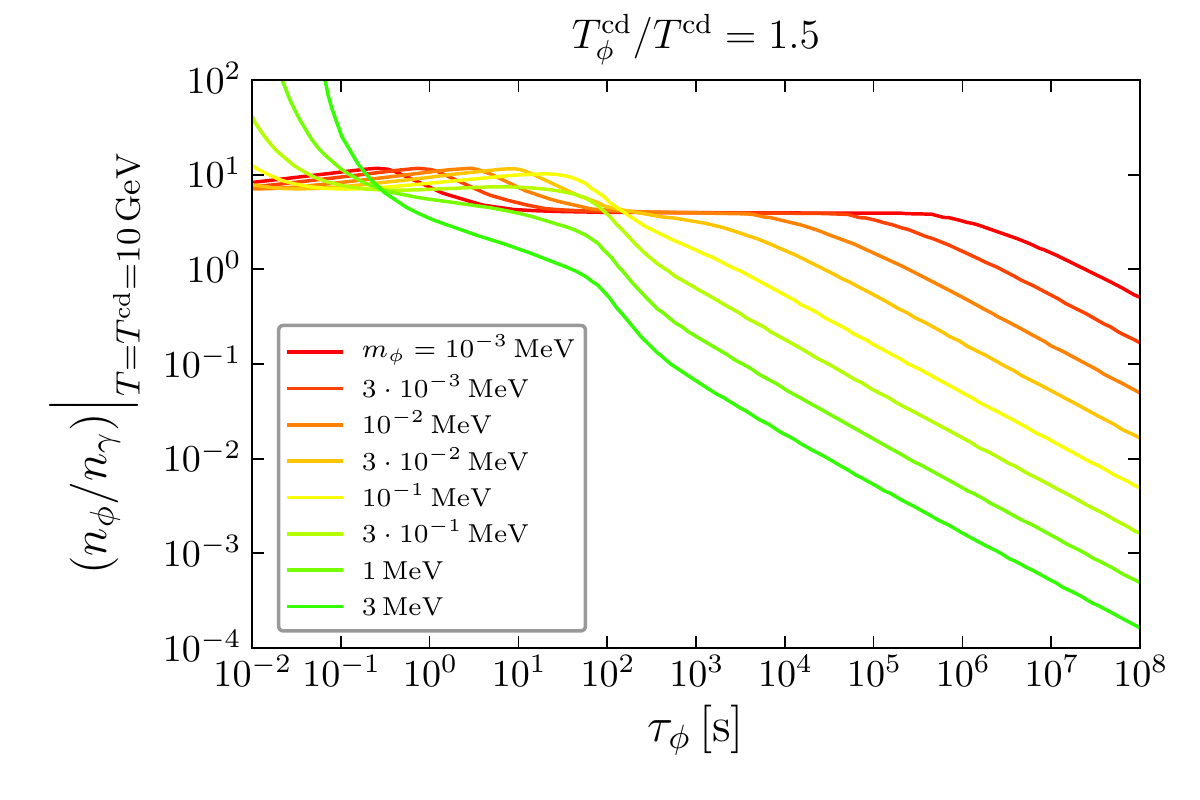}\hspace*{-0.2cm}
		\includegraphics[scale=0.68]{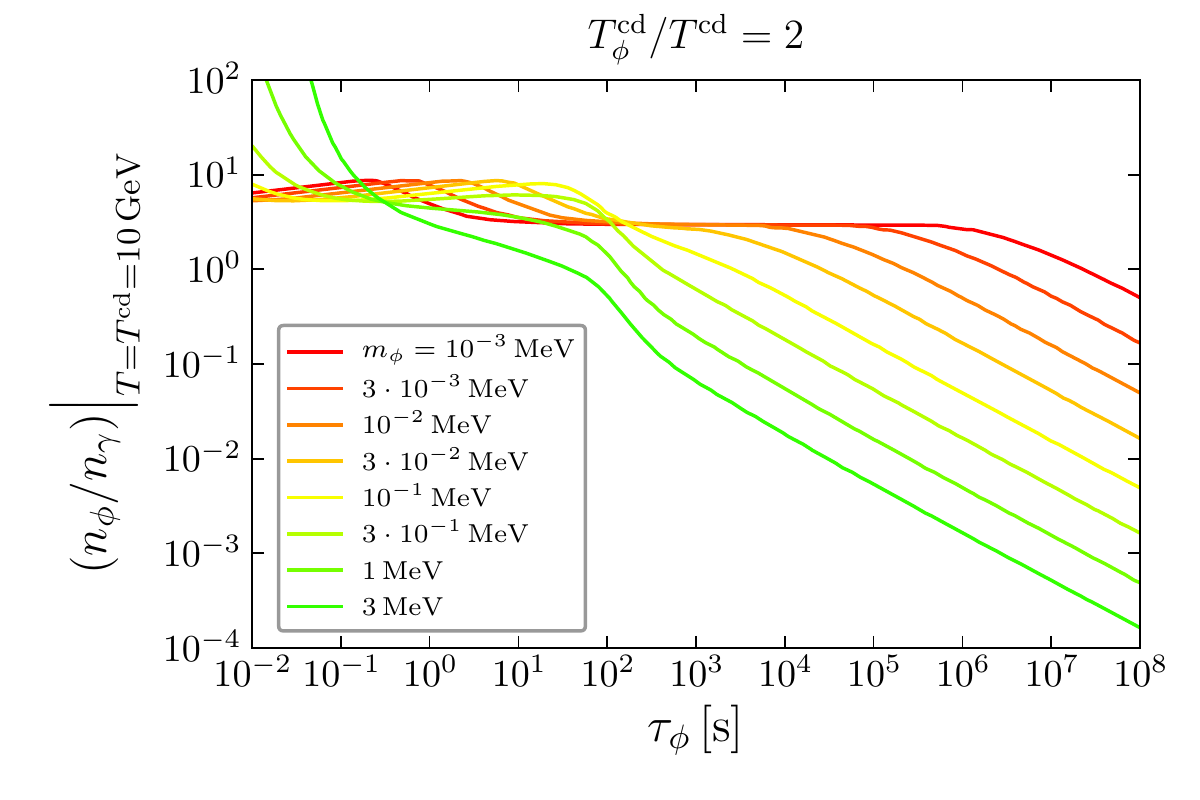}\\
		\hspace*{-0.7cm}
		\includegraphics[scale=0.68]{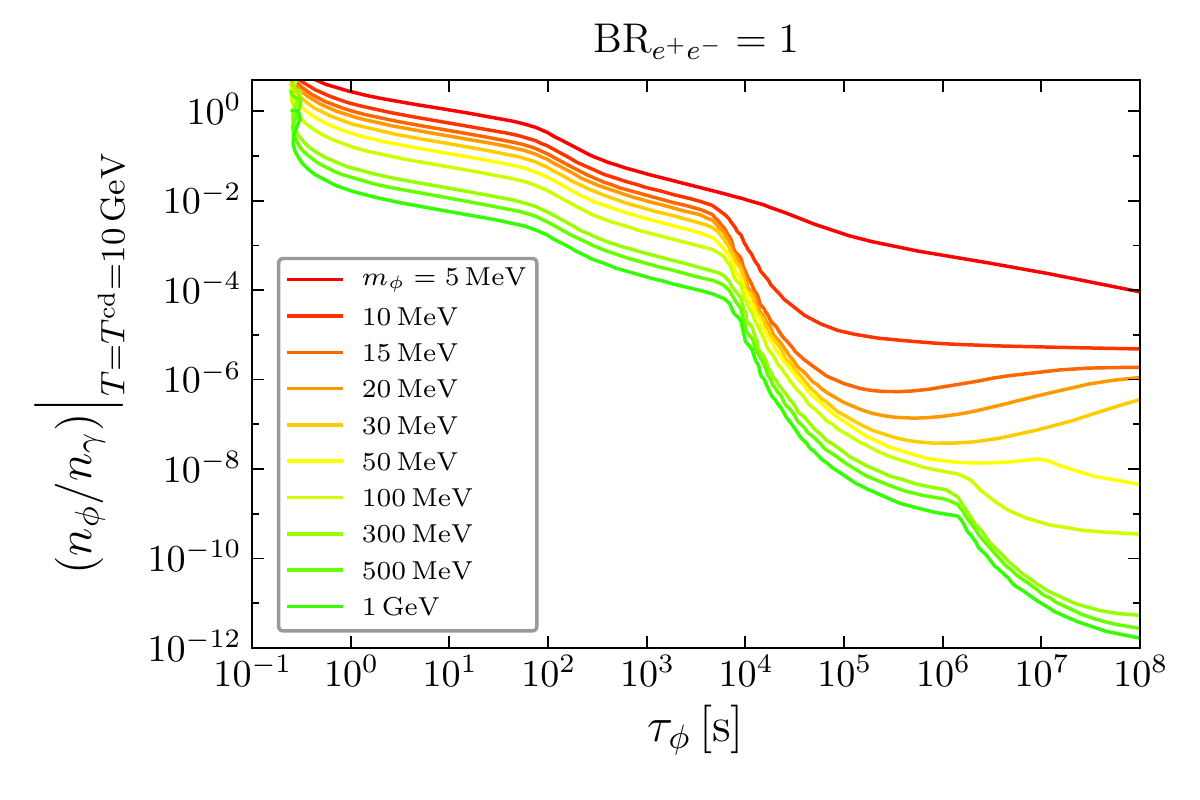}\hspace*{-0.2cm}
		\includegraphics[scale=0.68]{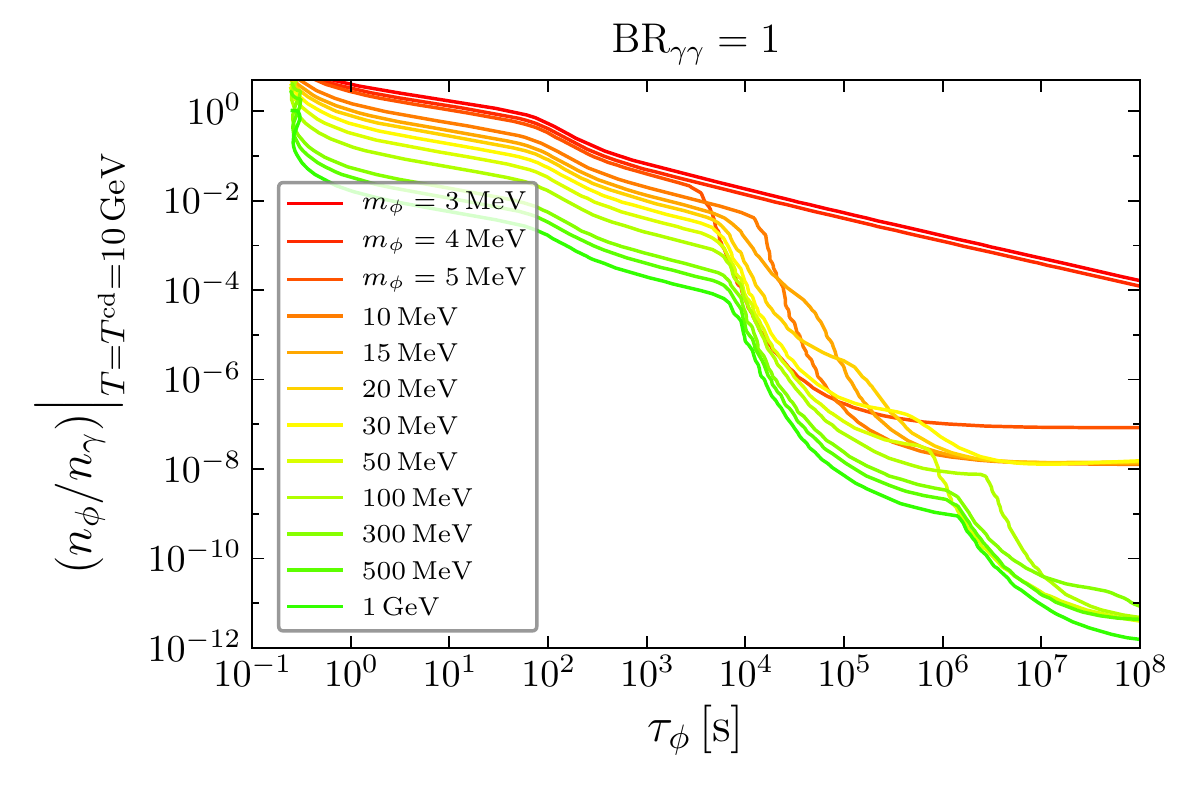}		
	\end{center}
	\caption{Upper limits from BBN on the initial abundance of $\phi$ relative to the photon number density. In the upper and central row we show the upper bounds for mediator masses below $3\,$MeV in which case the bounds are insensitive to the decay channel, but have a (mild) dependence on $T_\phi^\text{cd}/T^\text{cd}$, corresponding to the four different panels. In the lower row we present our results for various mediator masses $m_\phi \geq 5\,$MeV for decays into $e^+ e^-$ (left panel) and $\gamma \gamma$ (right panel), which in turn are independent of the ratio of the temperatures in the dark and visible sectors. See the text for more details on how to apply these bounds to a given model involving an unstable particle $\phi$.}
	\label{fig:appendix_plots}
\end{figure}

In figure~\ref{fig:appendix_plots} we provide upper limits from BBN on the abundance of $\phi$ as a function of its lifetime $\tau_\phi$, for a large number of masses $10^{-3}\,\text{MeV} \leq m_\phi \leq 1\,\text{GeV}$. Using this figure, it is straightforward to read off the BBN bound on a model involving an unstable particle with a given abundance prior to its decay. As follows from the discussion in section~\ref{sec:results}, for $m_\phi \lesssim 4\,$MeV the bound from BBN is insensitive to the decay channel (as photodisintegration is inactive), while in general it does depend on the temperature ratio $T_\phi^\text{cd}/T^\text{cd}$ at decoupling of $\phi$. Hence, we show the results for four different values $T_\phi^\text{cd}/T^\text{cd} = 0.5, 1, 1.5$ and $2$ in the upper and central row of figure~\ref{fig:appendix_plots}. Notice that the dependence on $T_\phi^\text{cd}/T^\text{cd}$ is only relevant for sufficiently small values of $\tau_\phi$ and if desired it can be easily interpolated to other values of the temperature ratio. On the other hand, for $m_\phi \gtrsim 3\,$MeV the bounds are insensitive to $T_\phi^\text{cd}/T^\text{cd}$ as the particle decays while being non-relativistic; however in that part of parameter space the bound can depend strongly on whether $\phi$ dominantly decays into $e^+ e^-$ or $\gamma \gamma$, corresponding to the left and right panel in the lower row of figure~\ref{fig:appendix_plots}, respectively.

In each panel, the vertical axis shows the ratio of the number density of $\phi$ prior to its decay relative to the photon number density, at a fixed reference temperature of $T^\text{cd} = 10\,$GeV after which $\phi$ by assumption is only subject to redshift and decay. If for the model of interest, the decoupling of $\phi$ happens at a different temperature $T^\text{cd}$, and possibly with a different dark sector temperature $T_\phi^\text{cd} \neq T^\text{cd}$, the BBN bound simply follows from rescaling the ones given in figure~\ref{fig:appendix_plots}, using the procedure outlined at the end of section~\ref{sec:upperlimits_abundance}.

\section{Rates for the cascade processes}
\label{app:rates_cascade}

In this appendix, we collect for reference all relevant total and differential interaction rates $\Gamma_X(E)$ and $K_{X' \to X}(E,E')$ for the cascade processes of high-energetic photons, electrons and positrons on the background photons, electrons and nuclei (see eqs.~(\ref{eq:fX_recursive}) and~(\ref{eq:FX_recursive})). 

\subsection*{Target densities}
The thermal photon spectrum $\bar{f}_\gamma(\be)$ is given by
\begin{equation}
\bar{f}_\gamma(\be) = \frac{\be^2}{\pi^2} \times \frac{1}{\exp(\be/T) - 1} \;\,,
\end{equation}
while the total baryon number density can be calculated from the baryon-to-photon ratio $\eta$ and the number density of photons $n_\gamma(T)$,
\begin{equation}
n_b(T) = \eta \times n_\gamma(T) = \eta \times \frac{2\zeta(3)}{\pi^2} T^3 \;\,.
\end{equation}
For the number density of background electrons $\bar{n}_e(T)$ we thus obtain
\begin{equation}
\bar{n}_e(T) = \sum_N Z_Nn_N \simeq (Y_{{}^1\text{H}} + 2\cdot Y_{{}^4\text{He}})\times n_b(T),\quad Y_N = \frac{n_N}{n_b} \;\,.
\end{equation}
At the times relevant to photodisintegration $(t \gtrsim 10^4\,\text{s})$, BBN has already terminated and the nuclear yields $Y_N$ are approximately constant. Hence, in the following we neglect the temperature dependence of $Y_N$, and fix them to their values directly after BBN, i.e.~to the values following from the calculation in section~\ref{sec:nuclabundance_calculation}.

\subsection*{Double photon pair creation: $\boldsymbol{\gamma + \gamma_\text{th} \rightarrow e^+ + e^-}$}
The rate for double photon pair creation is given by~\cite{Kawasaki:1994sc}\footnote{Correcting a typo in eq.~(27) of~\cite{Kawasaki:1994sc}.}
\begin{equation}
\Gamma_\gamma^\text{(DP)}(E) = \frac{1}{8E^2}\times \int_{m_e^2/E}^{\infty} \text{d}\be\; \frac{\bar{f}_\gamma(\be)}{\be^2} \times \int_{4m_e^2}^{4E\be} \text{d}s\;s\cdot\sigma_\text{DP}\left(\beta = \sqrt{1 - 4m_e^2/s}\right)
\end{equation}
with the total cross section
\begin{equation}
\sigma_\text{DP}(\beta) = \frac{\pi\alpha^2}{2 m_e^2} \times (1-\beta^2)\left[ (3-\beta^4)\ln\left( \frac{1+\beta}{1-\beta} \right) - 2\beta \left( 2 - \beta^2 \right) \right]\;\,.
\end{equation}
This process is only relevant above the threshold of production of electron-positron pairs $E \gtrsim m_e^2/(22T)$, allowing us to set $\Gamma_\gamma^\text{(DP)}(E) = 0$ for $E < m_e^2/(22T)$.

The differential rate for double photon pair creation entering the calculation of the electron and positron spectrum\footnote{Here the notation $\gamma \rightarrow e^\pm$ in the index of $K_{X' \rightarrow X}$ indicates that the corresponding expression is valid for $X' \rightarrow X \in \{\gamma \rightarrow e^+, \gamma \rightarrow e^-\}$ and consequently enters eq.~\eqref{eq:FX_recursive} twice.} was originally calculated in~\cite{1983Afz....19..323A} and is given by\footnote{Correcting a typo in eq.~(28) of~\cite{Kawasaki:1994sc}.}
\begin{equation}
K_{\gamma \rightarrow e^\pm}^\text{(DP)}(E, E') = \frac{\pi\alpha^2 m_e^2}{4} \times \frac{1}{E'^3}\int_{m_e^2/E'}^{\infty} \text{d}\be\;\frac{\bar{f}_\gamma(\be)}{\be^2}\;G(E, E', \be) \;\,,
\end{equation}
with
\begin{align}
G(E, E', \be) & = \frac{4(E' + \be)^2}{E(\Ep)}\ln\left( \frac{4\be E (\Ep)}{m_e^2(E' + \be)} \right) \nonumber \\ \nonumber \\
& + \left( \frac{m_e^2}{\be (E' + \be)} - 1 \right)\frac{(E'+\be)^4}{E^2(\Ep)^2} \nonumber \\ \nonumber \\
& + \frac{2\left[ 2\be (E' + \be) -m_e^2 \right](E'+\be)^2}{m_e^2 E (\Ep)} - 8\frac{\be (E'+\be)}{m_e^2}
\end{align}
for $m_e < E_\text{lim}^- < E < E_\text{lim}^+$,
\begin{equation}
2E_\text{lim}^\pm = E'+\be \pm (E'-\be)\sqrt{1- \frac{m_e^2}{E'\be}}\;\,,
\end{equation}
and $G(E, E', \be) = 0$ otherwise. As explained above, we furthermore set $K_{\gamma \rightarrow e^\pm}^\text{(DP)}(E, E') = 0$ for $E' < m_e^2/(22T)$.

\subsection*{Photon-photon scattering: $\boldsymbol{\gamma + \gamma_\text{th} \rightarrow \gamma + \gamma}$}
The total and differential interaction rate for photon-photon scattering have been originally calculated in~\cite{Svensson:1990pfo}, and are given by\footnote{Correcting a typo in eq.~(31) of~\cite{Kawasaki:1994sc} and in eq.~(5) of~\cite{Poulin:2015opa}.}
\begin{equation}
\Gamma_{\gamma}^{\text{(PP)}}(E) = \frac{1946}{50625\pi}\times \frac{8\pi^4}{63}\times \alpha^4 m_e \times \left( \frac{E}{m_e} \right)^3 \left( \frac{T}{m_e} \right)^6 \;\,,
\end{equation}
and
\begin{equation}
K_{\gamma \rightarrow \gamma}^\text{(PP)}(E, E') = \frac{1112}{10125\pi}\times\frac{\alpha^4}{m_e^8}\times \frac{8\pi^4T^6}{63} \times E'^2\left[ 1 - \frac{E}{E'} + \left( \frac{E}{E'} \right)^2 \right]^2 \;\,.
\end{equation}
In principle, these expressions are only valid for $E \lesssim m_e^2/T$~\cite{Kawasaki:1994sc}. However, for energies larger than this, photon-photon scattering is in any case negligible compared to double photon pair creation, making it unnecessary to impose this additional constraint.

\subsection*{Bethe-Heitler pair creation: $\boldsymbol{\gamma + N \rightarrow N + e^+ + e^-}$}
The total rate for Bethe-Heitler pair creation at energies $E \geq 4m_e$ and up to order $m_e^2/E^2$ can be written as~\cite{Maximon:1968, Kawasaki:1994sc}\footnote{We checked that higher order terms do not change the final results.}
\begin{eqnarray}
&&\Gamma_{\gamma}^{\text{(BH)}}(E) \simeq \frac{\alpha^3}{m_e^2} \times \Bigg( \sum_N Z_N^2 n_N(T) \Bigg) \times \Bigg( \left[ \frac{28}{9}\ln(2k) -\frac{218}{27} \right] \nonumber \\
&& \qquad+ \left(\frac{2}{k}\right)^2 \left[ \frac{2}{3}\ln(2k)^3 - \ln(2k)^2 + \left( 6 - \frac{\pi^2}{3} \right)\ln(2k) + 2\zeta(3) + \frac{\pi^2}{6} - \frac{7}{2}\right]\Bigg)\Bigg|_{k=E/m_e} \;\,.
\end{eqnarray}
Here, we only take into account scattering off $^1$H and $^4$He:
\begin{equation}
\sum_{N} Z_N^2 n_N(T) \simeq \sum_{N \in \{ {}^1\text{H}, {}^4\text{He} \}} Z_N^2 n_N(T) = (Y_{{}^1\text{H}} + 4 \cdot Y_{{}^4\text{He}}) \times n_b(T),\quad Y_N = \frac{n_N}{n_b}\;\,,
\end{equation}
since the abundances of all other nuclei are strongly suppressed. Furthermore, for energies in the range $2m_e < E \leq 4\;\mathrm{MeV}$, the interaction rate is essentially constant~\cite{Jedamzik:2006xz}: $\Gamma_{\gamma}^{\text{(BH)}}(E) \simeq \Gamma_{\gamma}^{\text{(BH)}}(E = 4\;\mathrm{MeV})$.

The differential rate for Bethe-Heitler pair creation is given by~\cite{berestetskii1982quantum, Kawasaki:1994sc}
\begin{align}
K_{\gamma \rightarrow e^\pm}^\text{(BH)}(E, E') = \Bigg( \sum_N Z_N^2 n_N(T) \Bigg) \times \frac{\text{d}\sigma_\text{BH}(E, E')}{\text{d}E} \times \Theta(E' - E-m_e) \,,
\label{kernel_BH_electron}
\end{align}
with the differential cross section
\begin{align}
\frac{\text{d}\sigma_\text{BH}(E, E')}{\text{d}E} = & \frac{\alpha^3}{m_e^2} \times \left( \frac{p_+ p_-}{E'^3} \right) \times \Bigg[ -\frac43 - 2E_+E_- \frac{p_+^2 + p_-^2}{p_+^2 p_-^2} \nonumber \\ \nonumber \\
& + m_e^2 \left( l_-\frac{E_+}{p_-^3} + l_+\frac{E_-}{p_+^3} - \frac{l_+ l_-}{p_+ p_-} \right) \nonumber \\ \nonumber \\
& + L\left( -\frac{8E_+E_-}{3p_+p_-} + \frac{E'^2}{p_+^3 p_-^3} \left( E_+^2E_-^2 + p_+^2 p_-^2 - m_e^2E_+ E_- \right) \right) \nonumber \\ \nonumber \\
& - L\frac{m_e^2E'}{2p_+ p_-}\left( l_+\frac{E_+ E_- - p_+^2}{p_+^3} + l_-\frac{E_- E_+ - p_-^2}{p_-^3} \right) \Bigg] \;\,,
\end{align}
where we have defined
\begin{align}
E_- \equiv E,\qquad E_+ \equiv E' - E&,\qquad p_\pm \equiv \sqrt{E_\pm^2 - m_e^2} \\ \nonumber \\
L \equiv \ln\left( \frac{E_+ E_- + p_+ p_- + m_e^2}{E_+ E_- - p_+ p_- + m_e^2} \right)&, \qquad l_\pm \equiv \ln\left( \frac{E_\pm + p_\pm}{E_\pm - p_\pm} \right) \;\,. \\ \nonumber
\end{align}
The $\Theta$-function appearing in eq.~\eqref{kernel_BH_electron} ensures that we fulfil energy conservation in the integration of $E'$ over the range $[E, \infty]$ in eq.~\eqref{eq:FX_recursive}.

\subsection*{Compton scattering: $\boldsymbol{\gamma + e^-_\text{th} \rightarrow \gamma + e^-}$}
The total rate for Compton scattering can be found in~\cite{Kawasaki:1994sc, Poulin:2015opa} and is given by
\begin{equation}
\Gamma_\gamma^\text{(CS)}(E) = \frac{2\pi\alpha^2}{m_e^2}\times\bar{n}_e(T)\times\frac{1}{x}\left[ \left(1 - \frac{4}{x} - \frac{8}{x^2}\right)\ln(1+x) + \frac12 + \frac8x - \frac{1}{2(1+x)^2}\right]\Bigg|_{x=2E/m_e} \;\,.
\end{equation}
Furthermore, the differential rate for the energy of the scattered photon reads~\cite{Kawasaki:1994sc, Poulin:2015opa}\footnote{Correcting a typo in eq.~(10) of~\cite{Poulin:2015opa}.}
\begin{align}
K_{\gamma \rightarrow \gamma}^\text{(CS)}(E, E') &= \Theta(E-E'/(1+2E'/m_e)) \times  \frac{\pi\alpha^2}{m_e}\times \bar{n}_e(T)\times \nonumber \\
&\frac{1}{E'^2}\left[ \frac{E'}{E} + \frac{E}{E'} + \left( \frac{m_e}{E} - \frac{m_e}{E'} \right)^2 - 2m_e\left( \frac{1}{E} - \frac{1}{E'} \right) \right]\;\,,
\label{kernel_CS_gamma}
\end{align}
with the $\Theta$-function corresponding to the vanishing of the rate above the Compton edge.

On the other hand, following~\cite{Kawasaki:1994sc}, the differential rate relevant for the spectrum of electrons can be deduced from eq.~\eqref{kernel_CS_gamma}:
\begin{equation}
K_{\gamma \rightarrow e^-}^\text{(CS)}(E, E') = K_{\gamma \rightarrow \gamma}^\text{(CS)}(E' + m_e - E, E')\;\,.
\end{equation}

\subsection*{Inverse Compton scattering: $\boldsymbol{e^\pm + \gamma_\text{th} \rightarrow e^\pm + \gamma}$}
The differential rate for production of photons from inverse Compton scattering was originally calculated in~\cite{Jones:1968zza} and can be written as
\begin{equation}
K_{e^\pm \rightarrow \gamma}^\text{(IC)}(E, E') = 2\pi\alpha^2\times \frac{1}{E'^2} \int_{0}^{\infty} \text{d}\be\;\frac{\bar{f}_\gamma(\be)}{\be}\;F(E, E', \be) \times \Theta(E' - E - m_e)\;\,.
\label{kernel_IC_gamma}
\end{equation}
For $\be \leq E \leq 4\be E'^2/(m_e^2 + 4\be E')$, the function $F(E, E', \be)$ is given by\footnote{Correcting a typo in eq.~(49) of~\cite{Kawasaki:1994sc}.}
\begin{equation}
F(E, E', \be) = 2q\ln(q) + (1+2q)(1-q)+\frac{\Gamma_\epsilon^2 q^2}{2+2\Gamma_\epsilon q}(1-q) \;\,,
\end{equation}
with
\begin{equation}
\Gamma_\epsilon = \frac{4\be E'}{m_e^2}, \qquad q = \frac{E}{\Gamma_\epsilon(E' - E)}\;\,,
\end{equation}
and $F(E, E', \be) = 0$ otherwise\footnote{According to~\cite{Jones:1968zza}, the function $F(E, E', \be)$ takes a different form for $E < \be$. However, this part of parameter space is practically irrelevant for our considerations.}. Again, the $\Theta$-function in eq.~\eqref{kernel_IC_gamma} ensures energy conservation upon integration of $E'$ over the range $[E, \infty]$.

The total rate for inverse Compton scattering entering the calculation of the electron and positron spectrum is given by~\cite{Jones:1968zza, Kawasaki:1994sc}\footnote{Correcting a typo in eq.~(48) of =~\cite{Kawasaki:1994sc}.}
\begin{equation}
\Gamma_{e^\pm}^\text{(IC)}(E) = 2\pi \alpha^2 \times \frac{1}{E^2}\int_{0}^{\infty} \text{d}E_\gamma\; \int_{0}^{\infty} \text{d}\be\; \frac{\bar{f}_\gamma(\be)}{\be} F(E_\gamma, E, \be) \;\,.
\end{equation}
Finally, the differential rate for the production of electrons and positrons can be written as~\cite{Jones:1968zza, Kawasaki:1994sc}
\begin{equation}
K_{e^\pm \rightarrow e^\pm}^\text{(IC)}(E, E') = 2\pi\alpha^2\times \frac{1}{E'^2} \int_{0}^{\infty} \text{d}\be\;\frac{\bar{f}_\gamma(\be)}{\be}\;F(E'+\be - E, E', \be) \;\,.
\end{equation}

\subsection*{Additional processes not considered in our calculation}
Other processes such as
\begin{itemize}
	\item Coulomb scattering $e^\pm/N + e^-_\text{th} \rightarrow e^\pm/N + e^-$\,,
	\item Thompson scattering $N + \gamma_\text{th} \rightarrow N + \gamma$\,,
	\item Magnetic moment scattering $N + e^-_\text{th} \rightarrow N + e^-$ or
	\item Electron-positron annihilation $e^+ + e^-_\text{th} \rightarrow \gamma + \gamma$
\end{itemize}
are suppressed by the small density of background electrons or nuclei  $\bar{n}_{e}, n_N \ll \bar{n}_{\gamma}$ and can therefore be neglected.

\bibliography{refs}

\providecommand{\bysame}{\leavevmode\hbox to3em{\hrulefill}\thinspace}
\begin{thebibliography}{10}

\bibitem{2016RvMP...88a5004C}
R.~H. {Cyburt}, B.~D. {Fields}, K.~A. {Olive}, and T.-H. {Yeh}, Reviews of
  Modern Physics \textbf{88} (2016), no.~1, 015004,  [1505.01076].

\bibitem{Olive:2016xmw}
Particle Data Group, C.~Patrignani et~al., Chin. Phys. \textbf{C40} (2016),
  no.~10, 100001.

\bibitem{Shvartsman:1969mm}
V.~F. Shvartsman, Pisma Zh. Eksp. Teor. Fiz. \textbf{9} (1969), 315--317, [JETP
  Lett.9,184(1969)].

\bibitem{Steigman:1977kc}
G.~Steigman, D.~N. Schramm, and J.~E. Gunn, Phys. Lett. \textbf{66B} (1977),
  202--204.

\bibitem{Scherrer:1987rr}
R.~J. Scherrer and M.~S. Turner, Astrophys. J. \textbf{331} (1988), 19--32,
  [Astrophys. J.331,33(1988)].

\bibitem{1988ApJ...331...33S}
R.~J. Scherrer and M.~S. Turner, Astrophysical Journal \textbf{331} (1988), 33.

\bibitem{Hufnagel:2017dgo}
M.~Hufnagel, K.~Schmidt-Hoberg, and S.~Wild, JCAP \textbf{1802} (2018), 044,
  [1712.03972].

\bibitem{Sarkar:1984tt}
S.~Sarkar and A.~M. Cooper-Sarkar, Phys. Lett. \textbf{148B} (1984), 347--354,
  [,I.362(1984)].

\bibitem{Ellis:1984er}
J.~R. Ellis, D.~V. Nanopoulos, and S.~Sarkar, Nucl. Phys. \textbf{B259} (1985),
  175--188.

\bibitem{Kawasaki:1994sc}
M.~Kawasaki and T.~Moroi, Astrophys. J. \textbf{452} (1995), 506,
  [astro-ph/9412055].

\bibitem{Cyburt:2002uv}
R.~H. Cyburt, J.~R. Ellis, B.~D. Fields, and K.~A. Olive, Phys. Rev.
  \textbf{D67} (2003), 103521,  [astro-ph/0211258].

\bibitem{Jedamzik:2006xz}
K.~Jedamzik, Phys. Rev. \textbf{D74} (2006), 103509,  [hep-ph/0604251].

\bibitem{Poulin:2015woa}
V.~Poulin and P.~D. Serpico, Phys. Rev. Lett. \textbf{114} (2015), no.~9,
  091101,  [1502.01250].

\bibitem{Poulin:2015opa}
V.~Poulin and P.~D. Serpico, Phys. Rev. \textbf{D91} (2015), no.~10, 103007,
  [1503.04852].

\bibitem{Arbey:2011nf}
A.~Arbey, Comput. Phys. Commun. \textbf{183} (2012), 1822--1831,  [1106.1363].

\bibitem{Buckley:2009in}
M.~R. Buckley and P.~J. Fox, Phys. Rev. \textbf{D81} (2010), 083522,
  [0911.3898].

\bibitem{Loeb:2010gj}
A.~Loeb and N.~Weiner, Phys. Rev. Lett. \textbf{106} (2011), 171302,
  [1011.6374].

\bibitem{Kaplinghat:2013yxa}
M.~Kaplinghat, S.~Tulin, and H.-B. Yu, Phys. Rev. \textbf{D89} (2014), no.~3,
  035009,  [1310.7945].

\bibitem{Kainulainen:2015sva}
K.~Kainulainen, K.~Tuominen, and V.~Vaskonen, Phys. Rev. \textbf{D93} (2016),
  no.~1, 015016,  [1507.04931].

\bibitem{Kahlhoefer:2017umn}
F.~Kahlhoefer, K.~Schmidt-Hoberg, and S.~Wild, JCAP \textbf{1708} (2017),
  no.~08, 003,  [1704.02149].

\bibitem{Kolb:1990vq}
E.~W. Kolb and M.~S. Turner, Front. Phys. \textbf{69} (1990), 1--547.

\bibitem{Chluba:2011hw}
J.~Chluba and R.~A. Sunyaev, Mon. Not. Roy. Astron. Soc. \textbf{419} (2012),
  1294--1314,  [1109.6552].

\bibitem{Poulin:2016anj}
V.~Poulin, J.~Lesgourgues, and P.~D. Serpico, JCAP \textbf{1703} (2017),
  no.~03, 043,  [1610.10051].

\bibitem{Dolgov:2002wy}
A.~D. Dolgov, Phys. Rept. \textbf{370} (2002), 333--535,  [hep-ph/0202122].

\bibitem{Kawasaki:1999na}
M.~Kawasaki, K.~Kohri, and N.~Sugiyama, Phys. Rev. Lett. \textbf{82} (1999),
  4168,  [astro-ph/9811437].

\bibitem{Fradette:2017sdd}
A.~Fradette and M.~Pospelov, Phys. Rev. \textbf{D96} (2017), no.~7, 075033,
  [1706.01920].

\bibitem{Aghanim:2018eyx}
Planck, N.~Aghanim et~al.,  (2018),  1807.06209.

\bibitem{Arbey:2018zfh}
A.~Arbey, J.~Auffinger, K.~P. Hickerson, and E.~S. Jenssen,  (2018),
  1806.11095.

\bibitem{Ade:2015xua}
Planck, P.~A.~R. Ade et~al., Astron. Astrophys. \textbf{594} (2016), A13,
  [1502.01589].

\bibitem{Forestell:2018txr}
L.~Forestell, D.~E. Morrissey, and G.~White,  (2018),  1809.01179.

\bibitem{Mardon:2009rc}
J.~Mardon, Y.~Nomura, D.~Stolarski, and J.~Thaler, JCAP \textbf{0905} (2009),
  016,  [0901.2926].

\bibitem{Birkedal:2005ep}
A.~Birkedal, K.~T. Matchev, M.~Perelstein, and A.~Spray,  (2005),
  hep-ph/0507194.

\bibitem{Bania:2002yj}
T.~M. Bania, R.~T. Rood, and D.~S. Balser, Nature \textbf{415} (2002), 54--57.

\bibitem{VangioniFlam:2002sa}
E.~Vangioni-Flam, K.~A. Olive, B.~D. Fields, and M.~Casse, Astrophys. J.
  \textbf{585} (2003), 611--616,  [astro-ph/0207583].

\bibitem{Kawasaki:2004qu}
M.~Kawasaki, K.~Kohri, and T.~Moroi, Phys. Rev. \textbf{D71} (2005), 083502,
  [astro-ph/0408426].

\bibitem{Fields:2011zzb}
B.~D. Fields, Ann. Rev. Nucl. Part. Sci. \textbf{61} (2011), 47--68,
  [1203.3551].

\bibitem{Korn:2006tv}
A.~J. Korn, F.~Grundahl, O.~Richard, P.~S. Barklem, L.~Mashonkina, R.~Collet,
  N.~Piskunov, and B.~Gustafsson, Nature \textbf{442} (2006), 657--659,
  [astro-ph/0608201].

\bibitem{Ma:2017ucp}
E.~Ma, Phys. Lett. \textbf{B772} (2017), 442--445,  [1704.04666].

\bibitem{Duerr:2018mbd}
M.~Duerr, K.~Schmidt-Hoberg, and S.~Wild,  (2018),  1804.10385.

\bibitem{Markevitch:2003at}
M.~Markevitch, A.~H. Gonzalez, D.~Clowe, A.~Vikhlinin, L.~David, W.~Forman,
  C.~Jones, S.~Murray, and W.~Tucker, Astrophys. J. \textbf{606} (2004),
  819--824,  [astro-ph/0309303].

\bibitem{Randall:2007ph}
S.~W. Randall, M.~Markevitch, D.~Clowe, A.~H. Gonzalez, and M.~Bradac,
  Astrophys. J. \textbf{679} (2008), 1173--1180,  [0704.0261].

\bibitem{Peter:2012jh}
A.~H.~G. Peter, M.~Rocha, J.~S. Bullock, and M.~Kaplinghat, Mon. Not. Roy.
  Astron. Soc. \textbf{430} (2013), 105,  [1208.3026].

\bibitem{Rocha:2012jg}
M.~Rocha, A.~H.~G. Peter, J.~S. Bullock, M.~Kaplinghat, S.~Garrison-Kimmel,
  J.~Onorbe, and L.~A. Moustakas, Mon. Not. Roy. Astron. Soc. \textbf{430}
  (2013), 81--104,  [1208.3025].

\bibitem{Kahlhoefer:2013dca}
F.~Kahlhoefer, K.~Schmidt-Hoberg, M.~T. Frandsen, and S.~Sarkar, Mon. Not. Roy.
  Astron. Soc. \textbf{437} (2014), no.~3, 2865--2881,  [1308.3419].

\bibitem{Harvey:2015hha}
D.~Harvey, R.~Massey, T.~Kitching, A.~Taylor, and E.~Tittley, Science
  \textbf{347} (2015), 1462--1465,  [1503.07675].

\bibitem{Kaplinghat:2015aga}
M.~Kaplinghat, S.~Tulin, and H.-B. Yu, Phys. Rev. Lett. \textbf{116} (2016),
  no.~4, 041302,  [1508.03339].

\bibitem{Bringmann:2016din}
T.~Bringmann, F.~Kahlhoefer, K.~Schmidt-Hoberg, and P.~Walia,  (2016),
  1612.00845.

\bibitem{Cirelli:2016rnw}
M.~Cirelli, P.~Panci, K.~Petraki, F.~Sala, and M.~Taoso, JCAP \textbf{1705}
  (2017), no.~05, 036,  [1612.07295].

\bibitem{Cline:2013gha}
J.~M. Cline, K.~Kainulainen, P.~Scott, and C.~Weniger, Phys. Rev. \textbf{D88}
  (2013), 055025,  [1306.4710], [Erratum: Phys. Rev.D92,no.3,039906(2015)].

\bibitem{Alekhin:2015byh}
S.~Alekhin et~al., Rept. Prog. Phys. \textbf{79} (2016), no.~12, 124201,
  [1504.04855].

\bibitem{Cassel:2009wt}
S.~Cassel, J. Phys. \textbf{G37} (2010), 105009,  [0903.5307].

\bibitem{Iengo:2009ni}
R.~Iengo, JHEP \textbf{05} (2009), 024,  [0902.0688].

\bibitem{Slatyer:2009vg}
T.~R. Slatyer, JCAP \textbf{1002} (2010), 028,  [0910.5713].

\bibitem{Steigman:2012nb}
G.~Steigman, B.~Dasgupta, and J.~F. Beacom, Phys. Rev. \textbf{D86} (2012),
  023506,  [1204.3622].

\bibitem{Krnjaic:2015mbs}
G.~Krnjaic, Phys. Rev. \textbf{D94} (2016), no.~7, 073009,  [1512.04119].

\bibitem{Raffelt:1987yu}
G.~G. Raffelt and D.~S.~P. Dearborn, Phys. Rev. \textbf{D36} (1987), 2211.

\bibitem{Aprile:2018dbl}
XENON, E.~Aprile et~al.,  (2018),  1805.12562.

\bibitem{Angloher:2015ewa}
CRESST, G.~Angloher et~al., Eur. Phys. J. \textbf{C76} (2016), no.~1, 25,
  [1509.01515].

\bibitem{Angloher:2017zkf}
CRESST, G.~Angloher et~al.,  (2017),  1701.08157.

\bibitem{Agnese:2015nto}
SuperCDMS, R.~Agnese et~al., Phys. Rev. Lett. \textbf{116} (2016), no.~7,
  071301,  [1509.02448].

\bibitem{Workgroup:2017lvb}
The GAMBIT Dark Matter Workgroup, T.~Bringmann et~al., Eur. Phys. J.
  \textbf{C77} (2017), no.~12, 831,  [1705.07920].

\bibitem{Athron:2018hpc}
GAMBIT, P.~Athron et~al.,  (2018),  1808.10465.

\bibitem{Angloher:2015eza}
CRESST, G.~Angloher et~al.,  (2015),  1503.08065.

\bibitem{Kahlhoefer:2017ddj}
F.~Kahlhoefer, S.~Kulkarni, and S.~Wild, JCAP \textbf{1711} (2017), no.~11,
  016,  [1707.08571].

\bibitem{Vogelsberger:2012ku}
M.~Vogelsberger, J.~Zavala, and A.~Loeb, Mon. Not. Roy. Astron. Soc.
  \textbf{423} (2012), 3740,  [1201.5892].

\bibitem{Tulin:2013teo}
S.~Tulin, H.-B. Yu, and K.~M. Zurek, Phys. Rev. \textbf{D87} (2013), no.~11,
  115007,  [1302.3898].

\bibitem{Tulin:2017ara}
S.~Tulin and H.-B. Yu,  (2017),  1705.02358.

\bibitem{Bezrukov:2009yw}
F.~Bezrukov and D.~Gorbunov, JHEP \textbf{05} (2010), 010,  [0912.0390].

\bibitem{Binder:2017lkj}
T.~Binder, M.~Gustafsson, A.~Kamada, S.~M.~R. Sandner, and M.~Wiesner, Phys.
  Rev. \textbf{D97} (2018), no.~12, 123004,  [1712.01246].

\bibitem{Bringmann:2018jpr}
T.~Bringmann, F.~Kahlhoefer, K.~Schmidt-Hoberg, and P.~Walia, Phys. Rev.
  \textbf{D98} (2018), no.~2, 023543,  [1803.03644].

\bibitem{1983Afz....19..323A}
F.~A. {Aharonian}, A.~M. {Atoian}, and A.~M. {Nagapetian}, Astrofizika
  \textbf{19} (1983), 323--334.

\bibitem{Svensson:1990pfo}
R.~Svensson and A.~A. Zdziarski, Astrophys. J. \textbf{349} (1990), 415--428.

\bibitem{Maximon:1968}
L.~C. Maximon, JOURNAL OF RESEARCH of the Notional Bureau of Standards (1968).

\bibitem{berestetskii1982quantum}
V.~Berestetskii, L.~Landau, E.~Lifshitz, L.~Pitaevskii, and J.~Sykes,
  \emph{Relativistic quantum theory}, 1971.

\bibitem{Jones:1968zza}
F.~C. Jones, Phys. Rev. \textbf{167} (1968), 1159--1169.

\end{thebibliography}
\bibliographystyle{ArXiv}

\end{document}